\documentclass[12pt]{iopart}

\usepackage{graphicx}

\begin{document}

\newcommand\Real{\mbox{Re}} 
\newcommand\Imag{\mbox{Im}} 
\newcommand\Rey{\mbox{\textit{Re}}} 
\newcommand\Pec{\mbox{\textit{Pe}}} 

\title[]{Numerical and experimental verification of a theoretical model of ripple formation 
in ice growth under supercooled water film flow}

\author{K Ueno$^1$\footnote{Corresponding author: k.ueno@kyudai.jp},
        M Farzaneh$^1$,
        S Yamaguchi$^2$ and
        H Tsuji$^3$} 

\address{$^1$NSERC/Hydro-Qu$\acute{\rm e}$bec/UQAC Industrial Chair on Atmospheric Icing of Power Network Equipment (CIGELE) and Canada Research Chair on Engineering of Power Network Atmospheric Icing (INGIVRE), Universit$\acute{\rm e}$ du Qu$\acute{\rm e}$bec $\grave{\rm a}$ Chicoutimi, 555 Boulevard de l'Universit$\acute{\rm e}$, Chicoutimi, Qu$\acute{\rm e}$bec, Canada, G7H 2B1 www.cigele.ca}
\address{$^2$Snow and Ice Research Center, National Research Institute for Earth Science and Disaster Prevention, Nagaoka, 940-0821, Japan}
\address{$^3$Research Institute for Applied Mechanics, Kyushu University, Kasuga, Fukuoka, 816-8580, Japan}


\begin{abstract}
Little is known about morphological instability of a solidification front during the crystal growth of a thin film of flowing supercooled liquid with a free surface: for example, the ring-like ripples on the surface of icicles. The length scale of the ripples is nearly 1 cm. Two theoretical models for the ripple formation mechanism have been proposed. However, these models lead to quite different results because of differences in the boundary conditions at the solid-liquid interface and liquid-air surface. The validity of the assumption used in the two models is numerically investigated and some of the theoretical predictions are compared with experiments. 
\end{abstract}

\vspace{2pc}

\maketitle

\section{Introduction \label{intro}}

\begin{figure}[t]
\begin{center}
\includegraphics[width=10cm,height=10cm,keepaspectratio,clip]{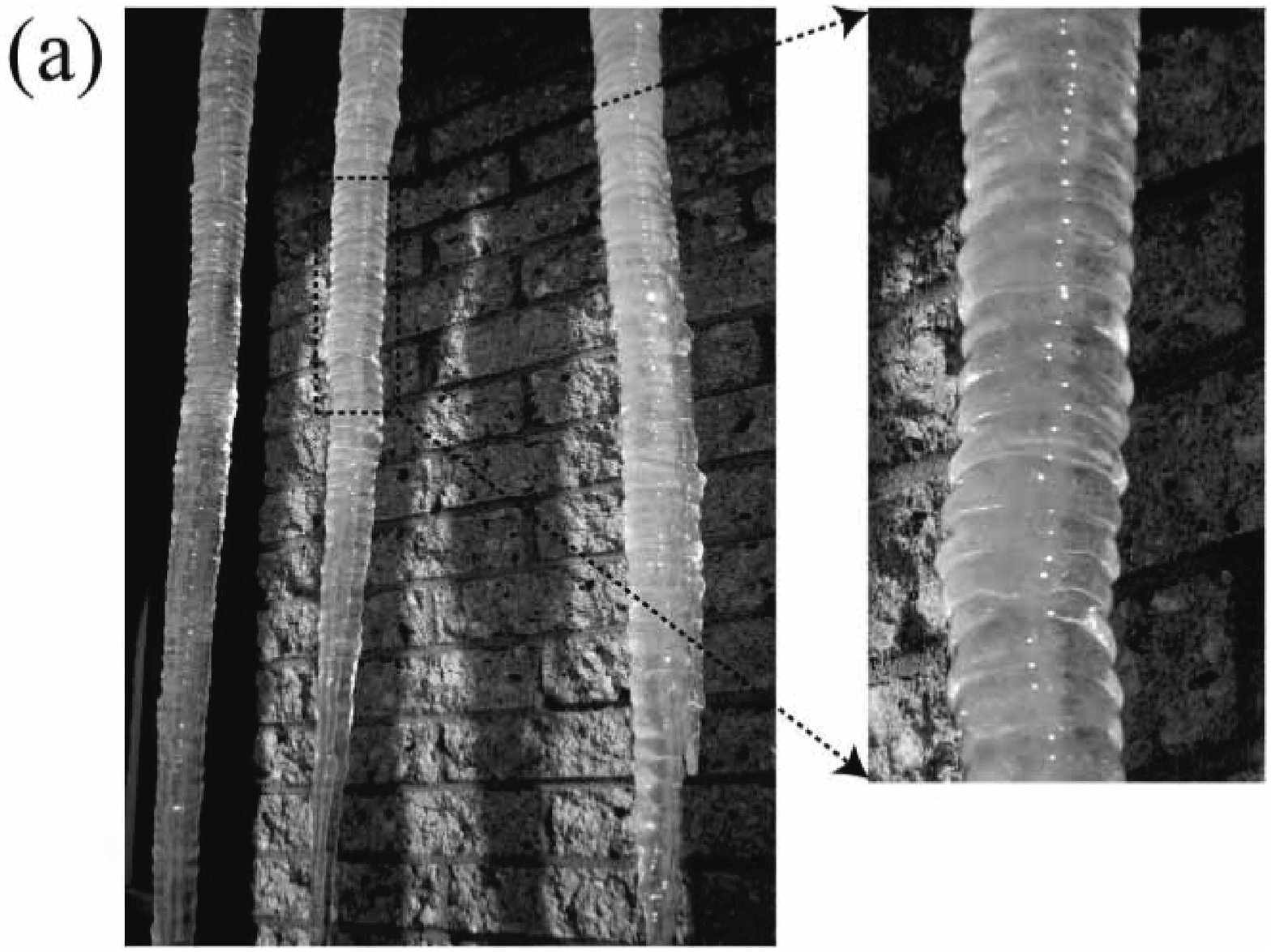}\\[0.5cm]
\includegraphics[width=7cm,height=7cm,keepaspectratio,clip]{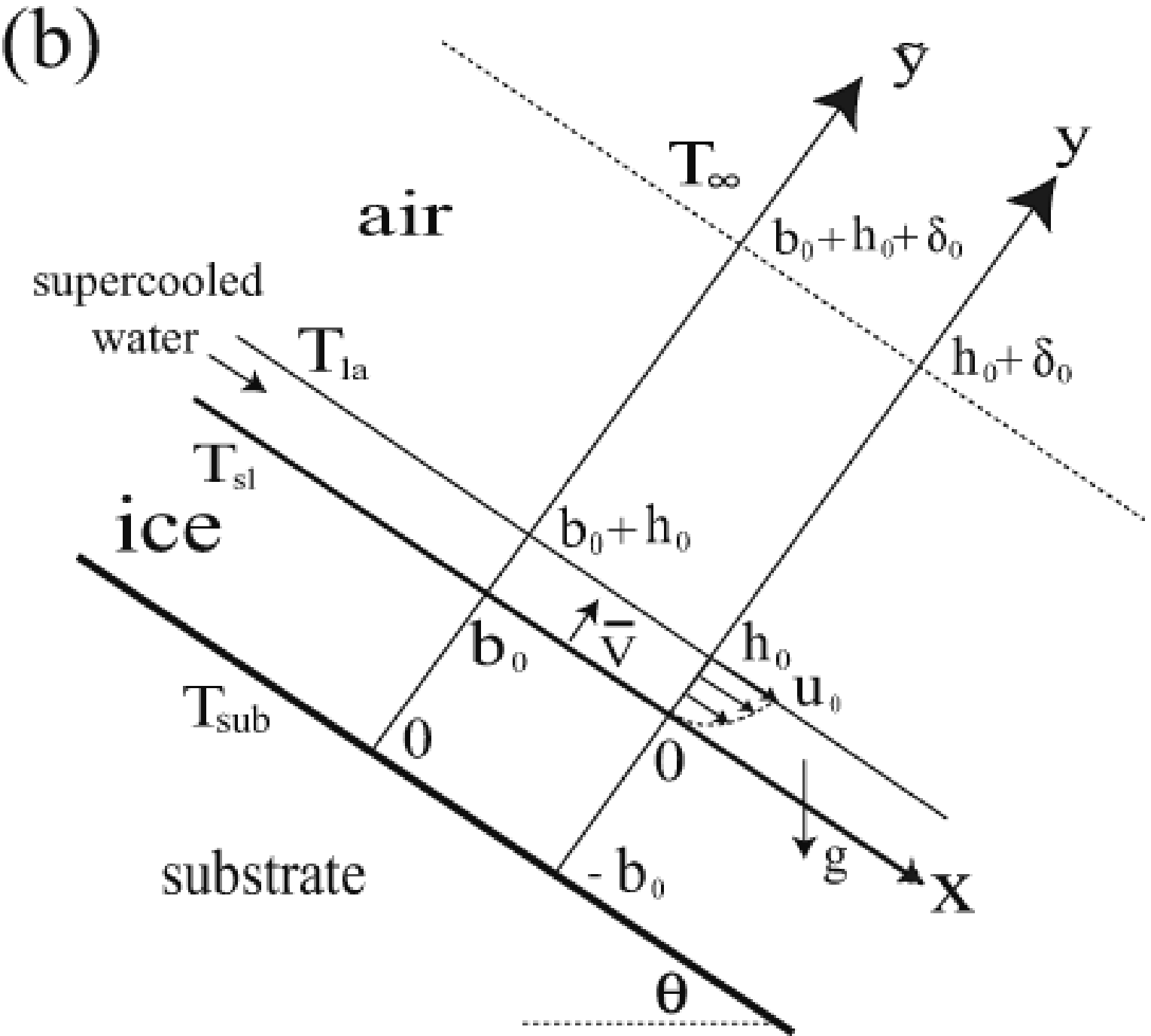}\hspace{5mm}
\includegraphics[width=7cm,height=7cm,keepaspectratio,clip]{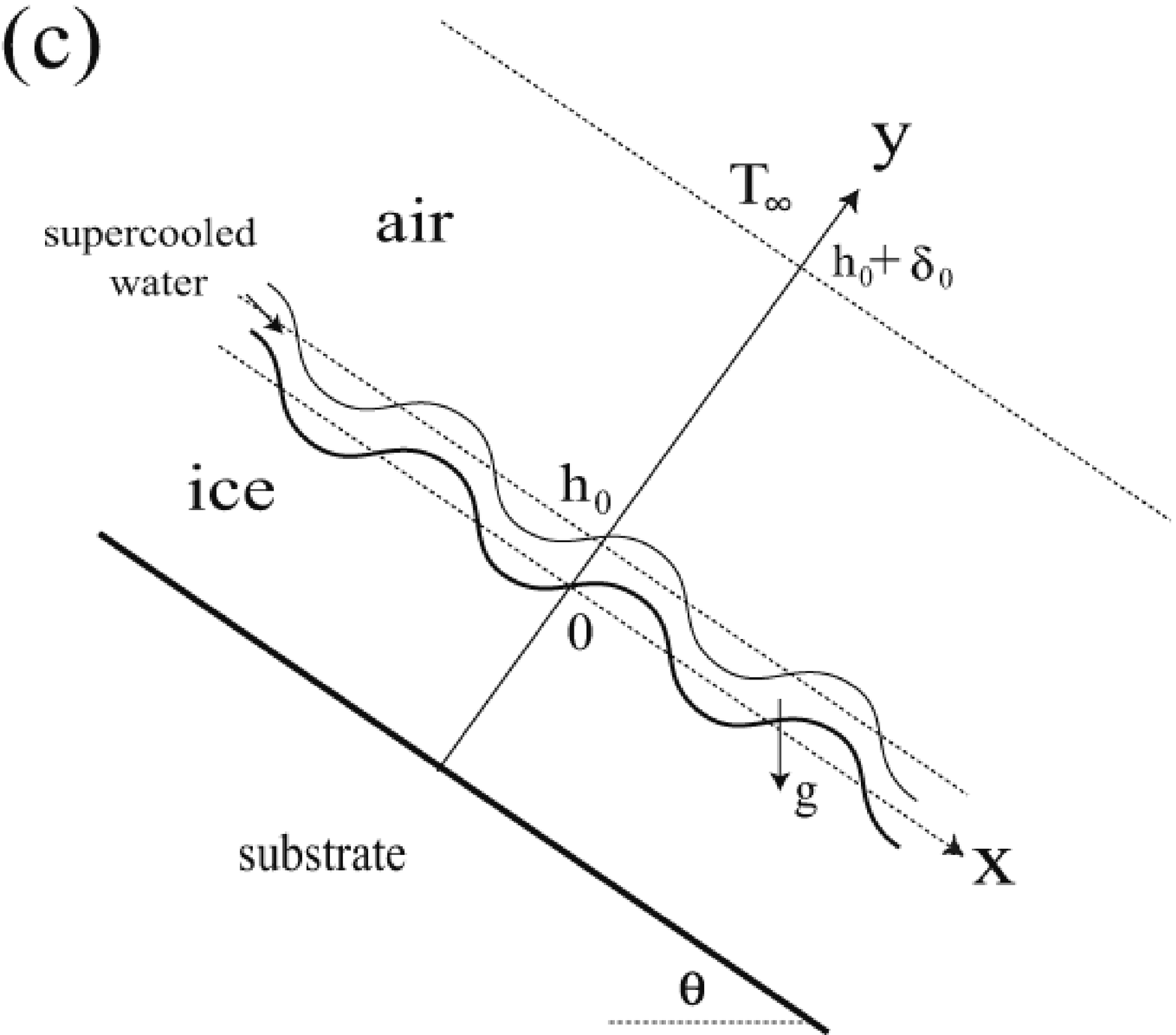}
\end{center}
\caption{(a) Ripples on natural icicles hanging from a roof (Feb 2009, Chicoutimi, Canada). (b) and (c) are schematic views of ice growth from a thin film of supercooled water flowing down a substrate inclined at angle $\theta$ to the 
horizontal. Flow is driven by gravity, and $g$ is the gravitational acceleration.  
(b) shows an unperturbed state of the ice-water interface and water-air surface, $h_{0}$ and $u_{0}$ are the thickness and the surface velocity of the flowing water film. 
$T_{\rm sub}$, $T_{sl}$, $T_{la}$ and $T_{\infty}$ are temperatures at the substrate ($y=-b_{0}$), ice-water interface ($y=0$) and water-air surface ($y=h_{0}$) and $y=h_{0}+\delta_{0}$, respectively.
The temperatures of the ice $T_{s}$, supercooled water $T_{l}$ and air $T_{a}$ are below 0 $^{\circ}$C.
$\bar{V}$ is an unperturbed ice growth rate. 
(c) shows a perturbed state of the ice-water interface and water-air surface.}
\label{fig:inclinedplane}
\end{figure}

A thin liquid film flowing down a rigid wall is often observed in everyday life.
A large number of studies on the instability of a viscous liquid layer running down a wall have been done (Benjamin 1957, Oron \etal 1997).
However, little is known about the morphological instability of the solid-liquid interface during the crystal growth of a flowing liquid film: for example, the ring-like ripples on icicles as shown in \fref{fig:inclinedplane} (a). 
Icicles grow when their surfaces are covered with a supercooled water film and the latent heat is released through the water film into the ambient air below 0 $^{\circ}$C (Makkonen 1988). It is well known that the ripples on the surface of icicles have a very regular spacing of about 9 mm (Maeno \etal 1994). A pattern similar to ripples on icicles was experimentally produced on the surface of a wooden round stick and that of a gutter on an inclined plane, by supplying water from their top in a cold room below 0 $^{\circ}$C  (Matsuda 1997). He found that these also have centimeter-scale ripples on their surfaces. 

What determines the length scale of ripples on the surface of icicles? Although this has been a familiar phenomenon for people in cold regions for a very long time (Terada 1947), nobody has been able to explain the details until recently. A first theoretical attempt to explain ripple formation on icicles was made in (Ogawa and Furukawa 2002). 
After that a quite different ripple formation mechanism was proposed by one of the authors (Ueno 2003, Ueno 2004, Ueno 2007).
Although the governing equations in both papers were basically identical, completely different results were obtained due to some differences in boundary conditions. The author (Ueno 2003, Ueno 2007) predicted that 
(i) the wavelength of ripples increases with a decrease in the angle of the inclined plane, 
(ii) the wavelength increases only gradually with an increase in the water supply rate per width, and  
(iii) the ripples move upward at about half speed of the mean growth rate of icicle radius. 
The experimental results by Matsuda were obtained at a fixed water supply rate. We conducted similar experiments for various inclination angles and water supply rates. The purpose of this paper is to numerically investigate the validity of the assumptions used in both models (Ogawa and Furukawa 2002) and (Ueno 2003), and to compare the above theoretical predictions (i), (ii) and (iii) with our own experimental results.

\section{Theoretical framework \label{frame}}

We consider an ice growth on a substrate by supplying water from the top, as shown in \fref{fig:inclinedplane} (b). 
One side of the water film is a water-air surface and the other side is growing ice. As a result of the instability of the ice-water interface as shown in \fref{fig:inclinedplane} (c), the flow in the water film can be changed depending on the morphology of ice. 
In previous papers (Ueno 2003, Ueno 2004, Ueno 2007), we assumed a semi-infinite ice layer and no airflow ahead of the water-air surface. 
In this paper, we extend the previous theoretical framework to include heat conduction from the ice-water interface into the substrate thorough the ice with finite thickness. Not only does this theoretical framework enable us to rewrite the governing equations and boundary conditions in a more tractable form to solve numerically, but it also let us  easily compare the difference between the models (Ogawa and Furukawa 2002) and (Ueno 2003).

Instead of dealing with the complete geometry of the icicle, round stick and gutter on an inclined plane, the theoretical analysis is assumed to be restricted to two-dimensional vertical cross-sections of their objects, as shown in figures \ref{fig:inclinedplane} (b) and (c). The $x$ axis is parallel to semi-parabolic shear flow direction, and the  $\tilde{y}$ and $y$ axes are normal to it. $\tilde{y}$ is a laboratory frame, and $y$ is a moving frame with an undisturbed ice growth rate $\bar{V}$. 
$h_{0}$ and $u_{0}$ are the thickness and the surface velocity of an undisturbed flowing supercooled water film. For typical values of water supply rate per width $Q/l$ (Maeno \etal 1994), $u_{0} \sim 1$ cm/s and $h_{0} \sim 100$ $\mu$m. Since actual thickness of the water film is very thin, the thickness of water film is drawn exaggeratedly in figures \ref{fig:inclinedplane} (b) and (c). For convenience, we list only non-standard or particularly important symbols in \Tref{Symbols}.

\begin{table}
\caption{Symbols.}
\begin{tabular}{@{}cl@{}}
\br
symbol & definition \\
\mr
$Q/l$                     & water supply rate per width \\
$\theta$                  & angle of inclined plane with respect to horizontal \\
$b_{0}$                   & thickness of growing ice \\
$h_{0}$                   & thickness of unperturbed water film  \\
$\delta_{0}$              & characteristic length scale of thickness of thermal boundary layer in air \\
$u_{0}$                   & surface velocity of flowing water film \\
$\zeta$                   & disturbed ice-water interface \\
$\xi$                     & disturbed water-air surface \\
$T_{\rm sub}$             & temperature of substrate \\
$T_{sl}$                  & temperature at ice-water interface \\
$T_{la}$                  & temperature at water-air surface \\
$T_{\infty}$              & ambient air temperature \\
$\Delta T_{sl}$           & temperature deviation from $T_{sl}$ \\
$\Delta T_{la}$           & temperature deviation from $T_{la}$ \\
$\Rey_{l}$                & Reynolds number of water film flow \\
$\Pec_{l}$                & Peclet number of water film flow \\
$f_{l}$                   & non-dimensional amplitude of perturbed stream function in water film \\
$H_{l}$                   & non-dimensional amplitude of perturbed temperature in water film \\
$\mu$                     & non-dimensional wave number \\
$\alpha$                  & restoring force due to surface tension and gravity \\
$\sigma^{(r)}_{*}$        & non-dimensional amplification rate of disturbed ice-water interface \\
$v_{p*}$                  & non-dimensional translational velocity of disturbed ice-water interface \\
$K^{s}_{l}$               & ratio of thermal conductivity of ice to that of water \\
$G^{s}_{l}$               & ratio of unperturbed temperature gradient in ice to that in water \\
$\Theta_{\xi_{*}}$        & phase shift of water-air surface against ice-water interface \\
$\Theta_{T_{\zeta_{*}}}$  & phase shift of temperature at ice-water interface against ice-water interface \\
$\Theta_{T_{\xi_{*}}}$    & phase shift of temperature at water-air surface against ice-water interface \\
$\Theta_{q_{l*}-q_{s*}}$  & phase shift of heat flux at ice-water interface against ice-water interface \\
\br
\end{tabular}
\label{Symbols}
\end{table}
 
\subsection{Governing equations \label{goveq}}

The velocity components $u_{l}$ and $v_{l}$ in the $x$ and $y$ directions in the water film flowing down an inclined plane at angle $\theta$ with respect to the horizontal are governed by the Navier-Stokes equations driven by gravity and the continuity equation (Landau and Lifshitz 1959):
\begin{equation}
\frac{\partial u_{l}}{\partial t}
+u_{l}\frac{\partial u_{l}}{\partial x}
+v_{l}\frac{\partial u_{l}}{\partial y} 
=-\frac{1}{\rho_{l}}\frac{\partial p_{l}}{\partial x}
+\nu_{l}\left(\frac{\partial^{2}u_{l}}{\partial x^{2}}
+\frac{\partial^{2}u_{l}}{\partial y^{2}}\right)+g\sin\theta,
\label{eq:gov-ul} 
\end{equation}

\begin{equation}
\frac{\partial v_{l}}{\partial t}
+u_{l}\frac{\partial v_{l}}{\partial x}
+v_{l}\frac{\partial v_{l}}{\partial y}
=-\frac{1}{\rho_{l}}\frac{\partial p_{l}}{\partial y}
+\nu_{l}\left(\frac{\partial^{2}v_{l}}{\partial x^{2}}
+\frac{\partial^{2}v_{l}}{\partial y^{2}}\right)-g\cos\theta, 
\label{eq:gov-vl}
\end{equation}

\begin{equation}
\frac{\partial u_{l}}{\partial x}+\frac{\partial v_{l}}{\partial y}=0,
\label{eq:conti}
\end{equation}
where $t$ is time, $p_{l}$ the pressure, $\rho_{l}=1.0 \times 10^{3}$ ${\rm kg/m^{3}}$, the density of water, $\nu_{l}=1.8\times 10^{-6}$ $\rm m^{2}/s$, the kinematic viscosity of water, g=9.8 $\rm m/s^{2}$, the gravitational acceleration. 
From (\ref{eq:conti}), using the stream function $\psi_{l}$, $u_{l}$ and $v_{l}$ can be expressed as  
$u_{l}=\partial \psi_{l}/\partial y$
and
$v_{l}=-\partial \psi_{l}/\partial x$.

The equations for the temperatures in the ice $T_{s}$, water $T_{l}$, and air $T_{a}$ are (Landau and Lifschitz 1959, Caroli \etal 1992)

\begin{equation}
\frac{\partial T_{s}}{\partial t}-\bar{V}\frac{\partial T_{s}}{\partial y}
=\kappa_{s}\left(\frac{\partial^{2} T_{s}}{\partial x^{2}}
+\frac{\partial^{2} T_{s}}{\partial y^{2}}\right),
\label{eq:gov-Ts}
\end{equation}

\begin{equation}
\frac{\partial T_{l}}{\partial t}+u_{l}\frac{\partial T_{l}}{\partial x}
+v_{l}\frac{\partial T_{l}}{\partial y}
=\kappa_{l}\left(\frac{\partial^{2} T_{l}}{\partial x^{2}}+\frac{\partial^{2} T_{l}}{\partial y^{2}}\right),
\label{eq:gov-Tl}
\end{equation}

\begin{equation}
\frac{\partial T_{a}}{\partial t}-\bar{V}\frac{\partial T_{a}}{\partial y}
=\kappa_{a}\left(\frac{\partial^{2} T_{a}}{\partial x^{2}}
+\frac{\partial^{2} T_{a}}{\partial y^{2}}\right),
\label{eq:gov-Ta}
\end{equation}
where $\kappa_{s}=1.15 \times 10^{-6}$ ${\rm m^{2}/s}$, $\kappa_{l}=1.33 \times 10^{-7}$ ${\rm m^{2}/s}$ and $\kappa_{a}=1.87 \times 10^{-5}$ ${\rm m^{2}/s}$ are the thermal diffusivities of the ice, water and air, respectively. We can neglect the second terms on the left hand side of (\ref{eq:gov-Ts}) and (\ref{eq:gov-Ta}) because ice grows very slowly (Ueno 2003).

\subsection{Boundary conditions \label{bc}}

\subsubsection{Hydrodynamic boundary conditions \label{ H-bc}}

Neglecting the density difference between ice and water,
both velocity components at a disturbed ice-water interface must satisfy (Myers \etal 2002a, Ogawa and Furukawa 2002)
\begin{equation}
u_{l}|_{y=\zeta}=0,
\hspace{1cm}
v_{l}|_{y=\zeta}=0.
\label{eq:Hb1}
\end{equation}
Except for (\ref{eq:Hb1}), the following boundary conditions are the same as those used in the stability analysis of a viscous liquid layer flowing down a rigid wall (Benjamin 1957).
The kinematic condition at a disturbed water-air surface is
\begin{equation}
\frac{\partial \xi}{\partial t}+u_{l}|_{y=\xi}\frac{\partial \xi}{\partial x}=v_{l}|_{y=\xi}.
\label{eq:Hb2}
\end{equation}
At the water-air surface the shear stress must vanish:
\begin{equation}
\frac{\partial u_{l}}{\partial y}\Big|_{y=\xi}
+\frac{\partial v_{l}}{\partial x}\Big|_{y=\xi}=0,
\label{eq:Hb3}
\end{equation}
and the normal stress including the stress induced by the surface tension $\gamma=7.6 \times 10^{-2}$ ${\rm N/m}$ of the water-air surface must balance the atmospheric pressure $P_{0}$:
\begin{equation}
-p_{l}|_{y=\xi}+2\rho_{l}\nu_{l}\frac{\partial v_{l}}{\partial y}\Big|_{y=\xi}
-\gamma\frac{\partial^{2}\xi}{\partial x^2}\left[1+\left(\frac{\partial \xi}{\partial x}\right)^{2}\right]^{-3/2}
=-P_{0}.
\label{eq:Hb4}
\end{equation}

\subsubsection{Thermodynamic boundary conditions \label{Th-bc}}

In the model (Ogawa and Furukawa 2002), the continuity of the temperature at a disturbed ice-water interface, $y=\zeta(t,x)$, is
\begin{equation}
T_{s}|_{y=\zeta}=T_{l}|_{y=\zeta}=T_{sl}.
\label{eq:Tb1-OF}
\end{equation}
If we can neglect the Gibbs-Thomson effect (temperature depression due to the curvature of the solid-liquid interface) (Caroli \etal 1992), $T_{sl}$ is assumed to be the equilibrium freezing temperature ($T_{sl}=0$ $^{\circ}$C for pure water).
On the other hand, in the model (Ueno 2003), the continuity condition is represented as follows:
\begin{equation}
T_{s}|_{y=\zeta}=T_{l}|_{y=\zeta}=T_{sl}+\Delta T_{sl}.
\label{eq:Tb1-Ueno}
\end{equation}
We will discuss in Section 4 that the temperature at a disturbed ice-water interface under a thin shear flow is not necessarily a constant $T_{sl}$, but that it deviates by $\Delta T_{sl}$ from  $T_{sl}$.
The heat conservation at the ice-water interface is (Langer 1980, Caroli \etal 1992, Ueno 2003)
\begin{equation}
L\left(\frac{\rmd {b}_{0}}{\rmd t}+\frac{\partial \zeta}{\partial t} \right)
=K_{s}\frac{\partial T_{s}}{\partial y}\Big|_{y=\zeta}
      -K_{l}\frac{\partial T_{l}}{\partial y}\Big|_{y=\zeta},
\label{eq:Tb2}
\end{equation}
where $L=3.3 \times 10^{8}$ ${\rm J/m^{3}}$ is the latent heat per unit volume, and $K_{s}=2.22$ $\rm J/(m\,K\,s)$ and $K_{l}=0.56$ $\rm J/(m\,K\,s)$ are the thermal conductivities of the ice and water, respectively. 

In the model (Ueno 2003), the continuity of the temperature at a disturbed water-air surface, $y=\xi(t,x)$, is
\begin{equation}
T_{l}|_{y=\xi}=T_{a}|_{y=\xi}=T_{la}.
\label{eq:Tb3-Ueno}
\end{equation}
We will discuss in Section 4 that the temperature at a disturbed water-air surface should remain at a constant $T_{la}$, which will be determined from the continuity of heat flux at the water-air surface. 
On the other hand, in the model (Ogawa and Furukawa 2002) the continuity of the temperature is
\begin{equation}
T_{l}|_{y=\xi}=T_{a}|_{y=\xi}=T_{la}+\Delta T_{la},
\label{eq:Tb3-OF}
\end{equation}
which means that the temperature at a disturbed water-air surface deviates by $\Delta T_{la}$ from $T_{la}$.
The heat conservation at the water-air surface is given by (Ogawa and Furukawa 2002, Ueno 2003)
\begin{equation}
-K_{l}\frac{\partial T_{l}}{\partial y}\Big|_{y=\xi}
=-K_{a}\frac{\partial T_{a}}{\partial y}\Big|_{y=\xi},
\label{eq:Tb4}
\end{equation}
where $K_{a}=0.024$ $\rm J/(m\,K\,s)$ is the thermal conductivity of the air. 

We will see later that completely different results in the two models arise from the different boundary conditions between (\ref{eq:Tb1-OF}), (\ref{eq:Tb3-OF}) in (Ogawa and Furukawa 2002) and (\ref{eq:Tb1-Ueno}), (\ref{eq:Tb3-Ueno}) in (Ueno 2003). $\Delta T_{sl}$ in (\ref{eq:Tb1-Ueno}) and $\Delta T_{la}$ in (\ref{eq:Tb3-OF}) cannot be determined {\it a priori}, but will be determined after solving the equation for the temperature in the flowing water film.

\subsection{Equations and solutions for unperturbed and perturbed fields \label{pert}}

Since a ring-like structure encircles the icicles and there is no noticeable azimuthal variation on the surface of the icicles (see \fref{fig:inclinedplane} (a)), it is sufficient to consider only a one dimensional perturbation in the $x$ direction of the ice-water interface,
$\zeta(t,x)=\zeta_{k}\exp[\sigma t+\rmi kx]$,
where $k$ is the wave number and $\sigma=\sigma^{(r)}+\rmi \sigma^{(i)}$, with $\sigma^{(r)}$ being the amplification rate and $v_{p} \equiv -\sigma^{(i)}/k$ being the phase velocity of the perturbation, and $\zeta_{k}$ is a small amplitude of the ice-water interface. 
We separate $\xi$, $\psi_{l}$, $p_{l}$, $T_{s}$, $T_{l}$ and $T_{a}$ into unperturbed steady fields and perturbed fields with prime as follows:
$\xi=h_{0}+\xi'$,
$\psi_{l}=\bar{\psi}_{l}+\psi'_{l}$,
$p_{l}=\bar{P}_{l}+p'_{l}$,
$T_{s}=\bar{T}_{s}+T'_{s}$,
$T_{l}=\bar{T}_{l}+T'_{l}$
and
$T_{a}=\bar{T}_{a}+T'_{a}$. We suppose that the respective perturbed parts are expressed as follows:
\begin{equation}
\left(
\begin{array}{c}
\xi'(t,x) \\ \psi'_{l}(t,x,y) \\ p'_{l}(t,x,y)\\ T'_{s}(t,x,y) \\ T'_{l}(t,x,y) \\ T'_{a}(t,x,y)  
\end{array}
\right)
=
\left(
\begin{array}{c}
\xi_{k} \\ F_{l}(y) \\ \Pi_{l}(y) \\ g_{s}(y) \\ g_{l}(y) \\ g_{a}(y) 
\end{array}
\right)
\exp[\sigma t+\rmi kx],
\label{eq:pertset}
\end{equation}
where $\xi_{k}$, $F_{l}$, $\Pi_{l}$, $g_{s}$, $g_{l}$ and $g_{a}$ are the amplitudes of respective perturbations and they are assumed to be of the order of $\zeta_{k}$. The calculation in the previous paper (Ueno 2003) was based on a linear stability analysis taking into account only the first order of $\zeta_{k}$. Furthermore, two approximations were used. The first is the long wavelength approximation (Benjamin 1957, Oron \etal 1997), which is valid when the water film thickness is much less than the characteristic length scale of ripples. Defining a dimensionless wave number by $\mu=kh_{0}$, we neglected the higher order of $\mu$. The second is the quasi-steady state approximation (Langer 1980, Caroli \etal 1992). We neglected the time derivative term of $u_{l}$, $v_{l}$, $T_{s}$, $T_{l}$, $T_{a}$, $\xi$ in (\ref{eq:gov-ul}), (\ref{eq:gov-vl}), (\ref{eq:gov-Ts}), (\ref{eq:gov-Tl}), (\ref{eq:gov-Ta}) and (\ref{eq:Hb2}) because these fields respond relatively rapidly to slow development of the ice-water interface perturbation. In order to check numerically the validity of the analytical results obtained under the long wavelength approximation in the previous papers (Ogawa and Furukawa 2002, Ueno 2003), we retain the higher order of $\mu$ in the following perturbed parts.

The equations of the unperturbed part in (\ref{eq:gov-ul}) and (\ref{eq:gov-vl}) are, respectively,
\begin{equation}
\nu_{l}\frac{\rmd^{2}\bar{U}_{l}}{\rmd y^{2}}+g\sin\theta=0, \qquad
-\frac{1}{\rho_{l}}\frac{\rmd\bar{P}_{l}}{\rmd y}-g\cos\theta=0.
\label{eq:gov-U-P}
\end{equation}
With the no-slip condition at the ice-water interface, $\bar{U}_{l}|_{y=0}=0$, the free shear stress at the water-air surface, $\rmd\bar{U}_{l}/\rmd y|_{y=h_{0}}=0$, and $\bar{P}_{l}|_{y=h_{0}}=P_{0}$, the solutions are
\begin{equation} 
\bar{U}_{l}(y)=u_{0}\left\{2\frac{y}{h_{0}}-\left(\frac{y}{h_{0}}\right)^{2}\right\}, \qquad
\bar{P}_{l}(y)=P_{0}-\rho_{l}g\cos\theta(y-h_{0}), 
\label{eq:sol-U-P}
\end{equation}
where $u_{0}=h_{0}^{2}g\sin\theta/(2\nu_{l})$ 
is the surface velocity of the water film. 
In the absence of ice growth, 
the water supply rate per width is given by $Q/l=\int_{0}^{h_{0}}\bar{U}_{l}(y)\rmd y=2u_{0}h_{0}/3$ in an undisturbed state (Benjamin 1957, Landau and Lifschitz 1959),
from which
$h_{0}$ and $u_{0}$ can be expressed with respect to experimentally controllable parameters $Q/l$ and $\theta$ as follows:
$h_{0}=[3\nu_{l}/(g\sin\theta)]^{1/3}(Q/l)^{1/3}$ and 
$u_{0}=[9g\sin\theta/(8\nu_{l})]^{1/3}(Q/l)^{2/3}$. 

From the dimensional consideration and the assumption that $F_{l}$ in (\ref{eq:pertset}) is of the order of $\zeta_{k}$, we assume $F_{l}(y)=u_{0}f_{l}(y)\zeta_{k}$. Substituting $F_{l}$ and $\Pi_{l}$ in (\ref{eq:pertset}) into the perturbed part of (\ref{eq:gov-ul}) and (\ref{eq:gov-vl}), and finally eliminating $\Pi_{l}$ from them by cross differentiation, we obtain the Orr-Sommerfeld equation for the non-dimensional amplitude of the perturbed stream function $f_{l}$ (Benjamin 1959):
\begin{equation}
\frac{\rmd^{4}f_{l}}{\rmd y_{*}^{4}}
=\left(2\mu^{2}+\rmi\mu \Rey_{l}\bar{U}_{l*}\right)\frac{\rmd^{2}f_{l}}{\rmd y_{*}^{2}}
-\left\{\mu^{4}+\rmi\mu \Rey_{l}\left(\mu^{2}\bar{U}_{l*}+\frac{\rmd^{2}\bar{U}_{l*}}{\rmd y_{*}^{2}}\right)\right\}f_{l},
\label{eq:gov-fl}
\end{equation}
where $y_{*}=y/h_{0}$, $\bar{U}_{l*}=\bar{U}_{l}/u_{0}=2y_{*}-y_{*}^{2}$ and $\Rey_{l}\equiv u_{0}h_{0}/\nu_{l}=3Q/(2l\nu_{l})$ is the Reynolds number. 

Linearizing the boundary conditions (\ref{eq:Hb1})-(\ref{eq:Hb4}) at the unperturbed ice-water interface $y=0$ and water air surface $y=h_{0}$, the perturbed part of the equations are, respectively, 
$\rmd\bar{U}_{l}/\rmd y|_{y=0}\zeta+u'_{l}|_{y=0}=0$,
$v'_{l}|_{y=0}=0$,
$\bar{U}_{l}|_{y=h_{0}}\partial \xi'/\partial x=v'_{l}|_{y=h_{0}}$,
$\rmd^{2}\bar{U}_{l}/\rmd y^{2}|_{y=h_{0}}\xi'+\partial u'_{l}/\partial y|_{y=h_{0}}+\partial v'_{l}/\partial x|_{y=h_{0}}=0$ and
$-(\rmd\bar{P}_{l}/\rmd y|_{y=h_{0}}\xi'+p'_{l}|_{y=h_{0}})+2\rho_{l}\nu_{l}\partial v'_{l}/\partial y|_{y=h_{0}}
-\gamma\partial^{2}\xi'/\partial x^{2}=0$.
Using $u'_{l}=\partial \psi'_{l}/\partial y$ and $v'_{l}=-\partial \psi'_{l}/\partial x$, the above equations 
can be expressed as respectively (Ueno 2003):
\begin{eqnarray}
\frac{\rmd f_{l}}{\rmd y_{*}}\Big|_{y_{*}=0}=-\frac{\rmd\bar{U}_{l*}}{\rmd y_{*}}\Big|_{y_{*}=0}, \qquad
f_{l}|_{y_{*}=0}=0, \qquad
f_{l}|_{y_{*}=1}\zeta_{k}=-\bar{U}_{l*}|_{y_{*}=1}\xi_{k}, \nonumber \\ 
\left(\frac{\rmd^{2}f_{l}}{\rmd y_{*}^{2}}\Big|_{y_{*}=1}+\mu^{2}f_{l}|_{y_{*}=1}\right)\zeta_{k}
=-\frac{\rmd^{2}\bar{U}_{l*}}{\rmd y_{*}^{2}}\Big|_{y_{*}=1}\xi_{k},\nonumber \\
\fl
\left\{\frac{\rmd^{3}f_{l}}{\rmd y_{*}^{3}}\Big|_{y_{*}=1}
-(\rmi\mu \Rey_{l}\bar{U}_{l*}|_{y_{*}=1}+3\mu^{2})\frac{\rmd f_{l}}{\rmd y_{*}}\Big|_{y_{*}=1}
+\rmi\mu \Rey_{l}\frac{\rmd\bar{U}_{l*}}{\rmd y_{*}}\Big|_{y_{*}=1}f_{l}|_{y_{*}=1}\right\}\zeta_{k}
=\rmi\alpha \xi_{k}.
\label{eq:original-bc-fl}
\end{eqnarray} 
While deriving the last equation in (\ref{eq:original-bc-fl}), we have used 
$\Pi_{l}=\rho_{l}u_{0}^{2}/h_{0}\{1/(\rmi \mu\Rey_{l})(\rmd^{3}f_{l}/\rmd y_{*}^{3}-\mu^{2}\rmd f_{l}/\rmd y_{*})
-\bar{U}_{l*}\rmd f_{l}/\rmd y_{*}+\rmd\bar{U}_{l*}/\rmd y_{*}f_{l}\}\zeta_{k}$ obtained from the perturbed part of pressure gradient term in (\ref{eq:gov-ul}). 
Here
\begin{equation}
\alpha=2(\cot\theta)\mu+\frac{2}{\sin\theta}\left(\frac{a}{h_{0}}\right)^{2}\mu^{3}
\label{eq:alpha}
\end{equation}
in the last equation in (\ref{eq:original-bc-fl}) is the parameter to characterize the effect of surface tension and gravity on the water-air surface, which was referred to as the restoring force in the papers (Benjamin 1957, Ueno 2003). $a=[\gamma/(\rho_{l}g)]^{1/2}$ is the capillary length associated with the surface tension $\gamma$ of the water-air surface (Landau and Lifschitz 1959). The third equation in (\ref{eq:original-bc-fl}) gives the relation between the amplitude of the ice-water interface, $\zeta_{k}$, and that of the water-air surface, $\xi_{k}$. 
In the case of a disturbed ice-water interface and water-air surface too, 
$\int_{\zeta}^{\xi}u_{l}(y)\rmd y
=\int_{\zeta}^{\xi}\{\bar{U}_{l}(y)+u'_{l}(x,y)\}\rmd y
=2u_{0}h_{0}/3+u_{0}(\xi_{k}+f_{l}|_{y_{*}=1}\zeta_{k})\exp[\sigma t+\rmi kx]$ up to the first order of $\zeta_{k}$ must be equal to $Q/l$, from which we again obtain the relation $\xi_{k}=-f_{l}|_{y_{*}=1}\zeta_{k}$.
Noting that 
$\bar{U}_{l*}|_{y_{*}=1}=1$,
$\rmd\bar{U}_{l*}/\rmd y_{*}|_{y_{*}=0}=2$ and 
$\rmd^{2}\bar{U}_{l*}/\rmd y_{*}^{2}|_{y_{*}=1}=-2$ 
and using $\xi_{k}=-f_{l}|_{y_{*}=1}\zeta_{k}$, (\ref{eq:original-bc-fl}) leads to four boundary conditions to solve (\ref{eq:gov-fl}):
\begin{eqnarray}
\frac{\rmd f_{l}}{\rmd y_{*}}\Big|_{y_{*}=0}+2=0,\qquad
f_{l}|_{y_{*}=0}=0, \qquad
\frac{\rmd^{2}f_{l}}{\rmd y_{*}^{2}}\Big|_{y_{*}=1}+(2+\mu^{2})f_{l}|_{y_{*}=1}=0,\nonumber \\
\frac{\rmd^{3}f_{l}}{\rmd y_{*}^{3}}\Big|_{y_{*}=1}
-(\rmi\mu \Rey_{l}+3\mu^{2})\frac{\rmd f_{l}}{\rmd y_{*}}\Big|_{y_{*}=1}+\rmi\alpha f_{l}|_{y_{*}=1}=0.
\label{eq:bc-fl}
\end{eqnarray} 

Linearizing (\ref{eq:Tb1-Ueno}) and (\ref{eq:Tb2}) at the unperturbed ice-water interface $y=0$ yields respectively, 
to the zeroth order in $\zeta_{k}$,
\begin{equation} 
\bar{T}_{s}|_{y=0}=\bar{T}_{l}|_{y=0}=T_{sl}, \hspace{5mm}
L\frac{\rmd b_{0}}{\rmd t}=K_{s}\frac{\rmd\bar{T}_{s}}{\rmd y}\Big|_{y=0}-K_{l}\frac{\rmd\bar{T}_{l}}{\rmd y}\Big|_{y=0},
\label{eq:bco(0)-sl-T}
\end{equation}  
and to the first order in $\zeta_{k}$,
\begin{eqnarray}
\fl
\Delta T_{sl}
=\left[\frac{\rmd\bar{T}_{s}}{\rmd y}\Big|_{y=0}\zeta_{k}+g_{s}|_{y=0}\right]\exp[\sigma t+\rmi kx]
=\left[\frac{\rmd\bar{T}_{l}}{\rmd y}\Big|_{y=0}\zeta_{k}+g_{l}|_{y=0}\right]\exp[\sigma t+\rmi kx],\nonumber \\
L\sigma\zeta_{k}=K_{s}\frac{\rmd g_{s}}{\rmd y}\Big|_{y=0}-K_{l}\frac{\rmd g_{l}}{\rmd y}\Big|_{y=0}.
\label{eq:bco(1)-sl-T}
\end{eqnarray} 
Next, linearizing (\ref{eq:Tb3-OF}) and (\ref{eq:Tb4}) at the unperturbed water-air surface $y=h_{0}$ yields respectively, 
to the zeroth order in $\xi_{k}$, 
\begin{equation} 
\bar{T}_{l}|_{y=h_{0}}=\bar{T}_{a}|_{y=h_{0}}=T_{la}, \hspace{1cm}
-K_{l}\frac{\rmd\bar{T}_{l}}{\rmd y}\Big|_{y=h_{0}}=-K_{a}\frac{\rmd\bar{T}_{a}}{\rmd y}\Big|_{y=h_{0}},
\label{eq:bco(0)-la-T}
\end{equation} 
and to the first order in $\xi_{k}$,
\begin{eqnarray}
\fl
\Delta T_{la}
=\left[\frac{\rmd\bar{T}_{l}}{\rmd y}\Big|_{y=h_{0}}\xi_{k}+g_{l}|_{y=h_{0}}\right]\exp[\sigma t+\rmi kx] 
=\left[\frac{\rmd\bar{T}_{a}}{\rmd y}\Big|_{y=h_{0}}\xi_{k}+g_{a}|_{y=h_{0}}\right]\exp[\sigma t+\rmi kx], \nonumber \\
K_{l}\left(\frac{\rmd^{2}\bar{T}_{l}}{\rmd y^{2}}\Big|_{y=h_{0}}\xi_{k}+\frac{\rmd g_{l}}{\rmd y}\Big|_{y=h_{0}}\right)
=K_{a}\left(\frac{\rmd^{2}\bar{T}_{a}}{\rmd y^{2}}\Big|_{y=h_{0}}\xi_{k}+\frac{\rmd g_{a}}{\rmd y}\Big|_{y=h_{0}}\right).
\label{eq:bco(1)-la-T}
\end{eqnarray}  

Substituting 
$T_{s}=\bar{T}_{s}+g_{s}\exp[\sigma t+\rmi kx]$, 
$T_{l}=\bar{T}_{l}+g_{l}\exp[\sigma t+\rmi kx]$ and 
$T_{a}=\bar{T}_{a}+g_{a}\exp[\sigma t+\rmi kx]$ into (\ref{eq:gov-Ts}), (\ref{eq:gov-Tl}) and (\ref{eq:gov-Ta}), the equations for $\bar{T}_{s}$, $\bar{T}_{l}$, $\bar{T}_{a}$ and $g_{s}$, $g_{l}$, $g_{a}$ are obtained.
With the following boundary conditions:  
$\bar{T}_{s}|_{y=-b_{0}}=T_{\rm sub}$, 
$\bar{T}_{s}|_{y=0}=\bar{T}_{l}|_{y=0}=T_{sl}$, 
$\bar{T}_{l}|_{y=h_{0}}=\bar{T}_{a}|_{y=h_{0}}=T_{la}$ and 
$\bar{T}_{a}|_{y=h_{0}+\delta_{0}}=T_{\infty}$, 
as shown in \fref{fig:inclinedplane} (b), 
we obtain linear temperature profiles
\begin{equation}
\fl
\bar{T}_{s}(y)=T_{\rm sub}+\bar{G}_{s}(y+b_{0}), \qquad
\bar{T}_{l}(y)=T_{sl}-\bar{G}_{l}y, \qquad 
\bar{T}_{a}(y)=T_{la}-\bar{G}_{a}(y-h_{0}),
\label{eq:sol-Tsla}
\end{equation}
where 
$\bar{G}_{s}=(T_{sl}-T_{\rm sub})/{b_{0}}$, 
$\bar{G}_{l}=(T_{sl}-T_{la})/{h_{0}}$ and
$\bar{G}_{a}=(T_{la}-T_{\infty})/{\delta_{0}}$ 
are the unperturbed part of temperature gradient. Here we have assumed that air temperature is approximately $T_{\infty}$ at a length scale $\delta_{0}$. In the presence of airflow, heat transport is greatly influenced by the convection, and $\delta_{0}$ is then regarded as a characteristic length scale of the thickness of the thermal boundary layer (Short \etal 2006). 

When we consider the icicle as a cylindrical object, the solutions of the unperturbed velocity and temperature profiles in the water film in the cylindrical coordinate under the assumption of axial symmetry with the same boundary conditions as the planar case are 
$\bar{U}_{l*}(r_{*})=R_{*}^{2}(1+1/R_{*})^{2}\ln(r_{*}/R_{*})-(r_{*}^{2}-R_{*}^{2})/2$ and
$\bar{T}_{l*}(r_{*})\equiv (\bar{T}_{l}-T_{sl})/(T_{sl}-T_{la})=-R_{*}\ln(r_{*}/R_{*})$,
where $r_{*}=r/h_{0}$ and $R_{*}=R/h_{0}$, $r$ and $R$ being the radial coordinate and the icicle radius, respectively. When we express $\bar{U}_{l*}(r_{*})$ and $\bar{T}_{l*}(r_{*})$ with respect to $y_{*}$ using the relation $r_{*}=R_{*}+y_{*}$, the planar velocity and temperature profiles $\bar{U}_{l*}(y_{*})=2y_{*}-y_{*}^2$ and $\bar{T}_{l*}(y_{*})=-y_{*}$ are retrieved because $y_{*}/R_{*} \ll 1$ in the water film ($0 \leq y_{*} \leq 1$) when the icicle radius $R$ is much greater than the thickness of water film $h_{0}$, i.e., $R_{*} \gg 1$. That is why the icicle geometry was approximated in the Cartesian coordinates.

Substituting the solutions of $\bar{T}_{l}$ and $\bar{T}_{a}$ in (\ref{eq:sol-Tsla}) into the second equation in (\ref{eq:bco(0)-la-T}), $T_{la}$ in (\ref{eq:Tb3-Ueno}) and (\ref{eq:Tb3-OF}) is obtained as follows:
\begin{equation}
T_{la}=\frac{T_{sl}+\frac{K_{a}}{K_{l}}\frac{h_{0}}{\delta_{0}}T_{\infty}}{1+\frac{K_{a}}{K_{l}}\frac{h_{0}}{\delta_{0}}}. 
\label{eq:Tla}
\end{equation}
Next substituting the solutions of $\bar{T}_{s}$ and $\bar{T}_{l}$  in (\ref{eq:sol-Tsla}) into the second equation in (\ref{eq:bco(0)-sl-T}) and using (\ref{eq:Tla}) yields an unperturbed ice growth rate approximately:
\begin{equation}
\bar{V}\equiv\frac{\rmd b_{0}}{\rmd t}=\frac{K_{s}}{L}\frac{T_{sl}-T_{\rm sub}}{b_{0}}+\frac{K_{a}}{L}\frac{T_{sl}-T_{\infty}}{\delta_{0}},
\label{eq:V}
\end{equation}
because $K_{a}/K_{l} \ll 1$ and $h_{0}/\delta_{0} \ll 1$.
In the presence of airflow, (\ref{eq:Tla}) and (\ref{eq:V}) are replaced by $T_{la}=\{T_{sl}+(K_{a}/K_{l})(h_{0}/\delta_{0})\bar{G}_{a*}T_{\infty}\}\{1+(K_{a}/K_{l})(h_{0}/\delta_{0})\bar{G}_{a*}\}$ and 
$\rmd b_{0}/\rmd t=(K_{s}/L)(T_{sl}-T_{\rm sub})/b_{0}+(K_{a}/L)(T_{sl}-T_{\infty})\bar{G}_{a*}/\delta_{0}$, where $\bar{G}_{a*}\equiv-\delta_{0}/(T_{la}-T_{\infty})\rmd\bar{T}_{a}/\rmd y|_{y=h_{0}}$
is the dimensionless air temperature gradient at the unperturbed water-air surface and  
depends on the Prandtle number of the air. In order to estimate the value of $\bar{G}_{a*}$, we need the exact temperature distribution $\bar{T}_{a}$ by solving the coupled Navier-Stokes and heat transport equations in the air (Short \etal 2006). As we will see later, however, as the air temperature gradient $\bar{G}_{a}$ at the water-air surface does not affect the wavelength of ripples on icicles, we assumed a linear air temperature profile ahead of the water-air surface, i.e., we consider the case of $\bar{G}_{a*}=1$. 

\begin{figure}[t]
\begin{center}
\includegraphics[width=7cm,height=7cm,keepaspectratio,clip]{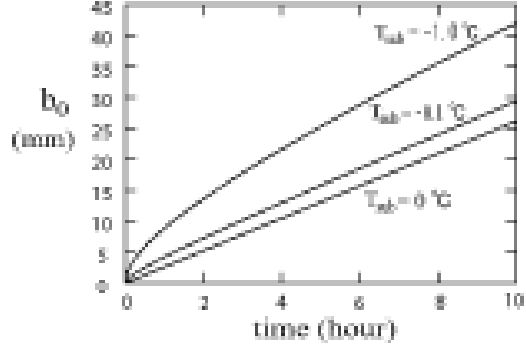}
\end{center}
\caption{Change in ice thickness with time for different temperatures $T_{\rm sub}$ of a substrate.
$\delta_{0}=1.0$ mm (Short \etal 2006), $T_{sl}=0$ $^{\circ}$C and $T_{\infty}=-10$ $^{\circ}$C.}
\label{fig:b0(t)}
\end{figure}

\Fref{fig:b0(t)} shows the ice thickness determined by integrating (\ref{eq:V}), subject to $b_{0}=0$ at time $t=0$, for different temperatures $T_{\rm sub}$ of a substrate. If heat conduction through the ice is negligible, $b_{0}$ is proportional to the time $t$ and $\bar{V}=2.6$ mm/h for the parameters shown in \fref{fig:b0(t)}. This value is of the same order as our experimental measurement of ice growth rate (see Section 5). When $T_{\rm sub} <0$ $^{\circ}$C, $b_{0}$ is proportional to $t^{1/2}$ because the first term on the right hand side in (\ref{eq:V}) is dominant while the thickness of ice is small.

From the dimensional consideration from the first equation in (\ref{eq:bco(1)-sl-T}), we assume $g_{l}(y)=H_{l}(y)\bar{G}_{l}\zeta_{k}$. Then the perturbed part of (\ref{eq:gov-Tl}) yields the equation for the non-dimensional amplitude $H_{l}$ (Ueno 2003): 
\begin{equation}
\frac{\rmd^{2}H_{l}}{\rmd y_{*}^{2}}
=(\mu^{2}+\rmi\mu \Pec_{l}\bar{U}_{l*})H_{l} 
-\rmi\mu \Pec_{l}\frac{\rmd\bar{T}_{l*}}{\rmd y_{*}}f_{l},
\label{eq:gov-Hl}
\end{equation}
where $\bar{T}_{l*}(y_{*})\equiv (\bar{T}_{l}(y_{*})-T_{sl})/(T_{sl}-T_{la})=-y_{*}$ is the dimensionless temperature profile of the water film in the unperturbed state, and where $\Pec_{l}\equiv u_{0}h_{0}/\kappa_{l}=3Q/(2l\kappa_{l})$ is the Peclet number defined as the ratio of the heat transfer due to the water flow to that due to the thermal diffusion in the water film.
 
In the case of $\Delta T_{la}=0$ as in (\ref{eq:Tb3-Ueno}) (Ueno 2003), the first equation in (\ref{eq:bco(1)-la-T}) gives 
$H_{l}|_{y=h_{0}}\zeta_{k}=\xi_{k}$ and 
$g_{a}|_{y=h_{0}}=\bar{G}_{a}\xi_{k}$.
Then the solution of the equation $\rmd^{2}g_{a}/\rmd y^{2}=k^{2}g_{a}$ with the boundary conditions of
$g_{a}|_{y=h_{0}}=\bar{G}_{a}\xi_{k}$ and $g_{a}|_{y=\infty}=0$ is given by
\begin{equation}
g_{a}(y)=\exp[-k(y-h_{0})]\bar{G}_{a}\xi_{k}.
\label{eq:sol-ga-Ueno}
\end{equation}
Substituting (\ref{eq:sol-ga-Ueno}) into the second equation in (\ref{eq:bco(1)-la-T}) 
and using the second equation $K_{l}\bar{G}_{l}=K_{a}\bar{G}_{a}$ in (\ref{eq:bco(0)-la-T}), yields
$\rmd H_{l}/\rmd y|_{y=h_{0}}\zeta_{k}=-k\xi_{k}$.
Accordingly, using the relation $\xi_{k}=-f_{l}|_{y_{*}=1}\zeta_{k}$ the boundary conditions to solve (\ref{eq:gov-Hl}) are   
\begin{equation}
H_{l}|_{y_{*}=1}=-f_{l}|_{y_{*}=1}, \qquad
-\frac{\rmd H_{l}}{\rmd y_{*}}\Big|_{y_{*}=1}=-\mu f_{l}|_{y_{*}=1}.
\label{eq:bc-Hl-Ueno}
\end{equation}
Using the solution $H_{l}$ satisfying the boundary conditions in (\ref{eq:bc-Hl-Ueno}), from the first equation in (\ref{eq:bco(1)-sl-T}), $\Delta T_{sl}$ and $g_{s}|_{y=0}$ are finally determined to the first order in $\zeta_{k}$: 
$\Delta T_{sl}=(H_{l}|_{y_{*}=0}-1)\bar{G}_{l}\zeta_{k}\exp[\sigma t+\rmi kx]$ and $g_{s}|_{y=0}=(-\bar{G}_{s}/\bar{G}_{l}+H_{l}|_{y=0}-1)\bar{G}_{l}\zeta_{k}$, respectively.
The solution to the equation $\rmd^{2}g_{s}/\rmd y^{2}=k^{2}g_{s}$ with the boundary conditions of 
$g_{s}|_{y=-b_{0}}=0$ and $g_{s}|_{y=0}=(-\bar{G}_{s}/\bar{G}_{l}+H_{l}|_{y=0}-1)\bar{G}_{l}\zeta_{k}$ is
\begin{equation}
g_{s}(y)=\left(-\frac{\bar{G}_{s}}{\bar{G}_{l}}+H_{l}|_{y=0}-1\right)\frac{\sinh[k(y+b_{0})]}{\sinh(kb_{0})}\bar{G}_{l}\zeta_{k}.
\label{eq:sol-gs-Ueno}
\end{equation}

On the other hand, in the case of $\Delta T_{sl}=0$ as in (\ref{eq:Tb1-OF}) (Ogawa and Furukawa 2002), the first equation in (\ref{eq:bco(1)-sl-T}) gives 
$g_{s}|_{y=0}=-\bar{G}_{s}\zeta_{k}$ and 
$H_{l}|_{y=0}=1$.
Then the solution to the equation $\rmd^{2}g_{s}/\rmd y^{2}=k^{2}g_{s}$ with the boundary conditions of 
$g_{s}|_{y=-b_{0}}=0$ and $g_{s}|_{y=0}=-\bar{G}_{s}\zeta_{k}$ is given by 
\begin{equation}
g_{s}(y)=-\frac{\sinh[k(y+b_{0})]}{\sinh(kb_{0})}\bar{G}_{s}\zeta_{k}.
\label{eq:sol-gs-OF}
\end{equation}
Furthermore, the first equation in (\ref{eq:bco(1)-la-T}) gives 
$g_{a}|_{y=h_{0}}=\bar{G}_{a}\xi_{k}+\bar{G}_{l}(H_{l}|_{y=h_{0}}\zeta_{k}-\xi_{k})$.
Then the solution to the equation $\rmd^{2}g_{a}/\rmd y^{2}=k^{2}g_{a}$ with the boundary conditions of
$g_{a}|_{y=h_{0}}=\bar{G}_{a}\xi_{k}+\bar{G}_{l}(H_{l}|_{y=h_{0}}\zeta_{k}-\xi_{k})$ and $g_{a}|_{y=\infty}=0$ takes the form:
\begin{equation}
g_{a}(y)=\exp[-k(y-h_{0})]\{\bar{G}_{a}\xi_{k}+\bar{G}_{l}(H_{l}|_{y=h_{0}}\zeta_{k}-\xi_{k})\}, 
\label{eq:sol-ga-OF}
\end{equation}
where $H_{l}$ is not yet determined.  
Substituting (\ref{eq:sol-ga-OF}) into the second equation in (\ref{eq:bco(1)-la-T}) yields
$\rmd H_{l}/\rmd y|_{y=h_{0}}K_{l}\bar{G}_{l}\zeta_{k}=-kK_{a}\{\bar{G}_{a}\xi_{k}+\bar{G}_{l}(H_{l}|_{y=h_{0}}\zeta_{k}-\xi_{k})\}$.
Then, using the relation $\xi_{k}=-f_{l}|_{y_{*}=1}\zeta_{k}$ the boundary conditions to solve (\ref{eq:gov-Hl}) are  
\begin{equation}
\fl
H_{l}|_{y_{*}=0}=1, \qquad
-\frac{\rmd H_{l}}{\rmd y_{*}}\Big|_{y_{*}=1}
=-\mu\left\{f_{l}|_{y_{*}=1}
-\frac{K_{a}}{K_{l}}(H_{l}|_{y_{*}=1}+f_{l}|_{y_{*}=1})\right\}.
\label{eq:bc-Hl-OF}
\end{equation}
Since $K_{a}/K_{l} \ll 1$, we can neglect the second term of the second equation in (\ref{eq:bc-Hl-OF}) and the result is the same as the second equation in (\ref{eq:bc-Hl-Ueno}).
Finally, substituting the solution of $H_{l}$ satisfying the boundary conditions in (\ref{eq:bc-Hl-OF}) into (\ref{eq:sol-ga-OF}), the solution of $g_{a}$ is determined and $\Delta T_{la}$ is obtained from the first equation in (\ref{eq:bco(1)-la-T}) to the first order in $\zeta_{k}$: 
$\Delta T_{la}=(H_{l}|_{y_{*}=1}+f_{l}|_{y_{*}=1})\bar{G}_{l}\zeta_{k}\exp[\sigma t+\rmi kx]$.
We will see later that the difference in the boundary conditions between (\ref{eq:bc-Hl-Ueno}) (Ueno 2003) and (\ref{eq:bc-Hl-OF}) (Ogawa and Furukawa 2002) is the main cause leading to different results.

\begin{figure}[t]
\begin{center}
\includegraphics[width=7.5cm,height=7.5cm,keepaspectratio,clip]{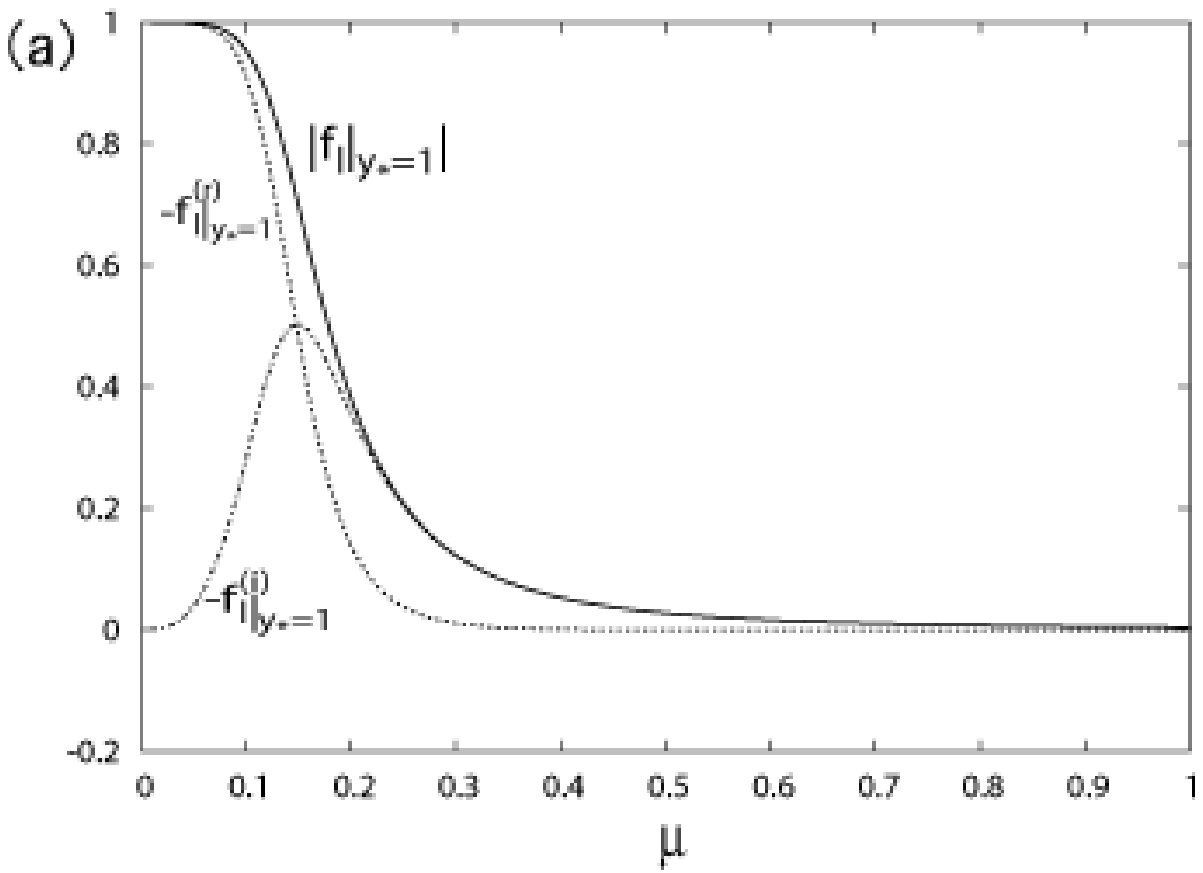}\hspace{5mm}
\includegraphics[width=7.5cm,height=7.5cm,keepaspectratio,clip]{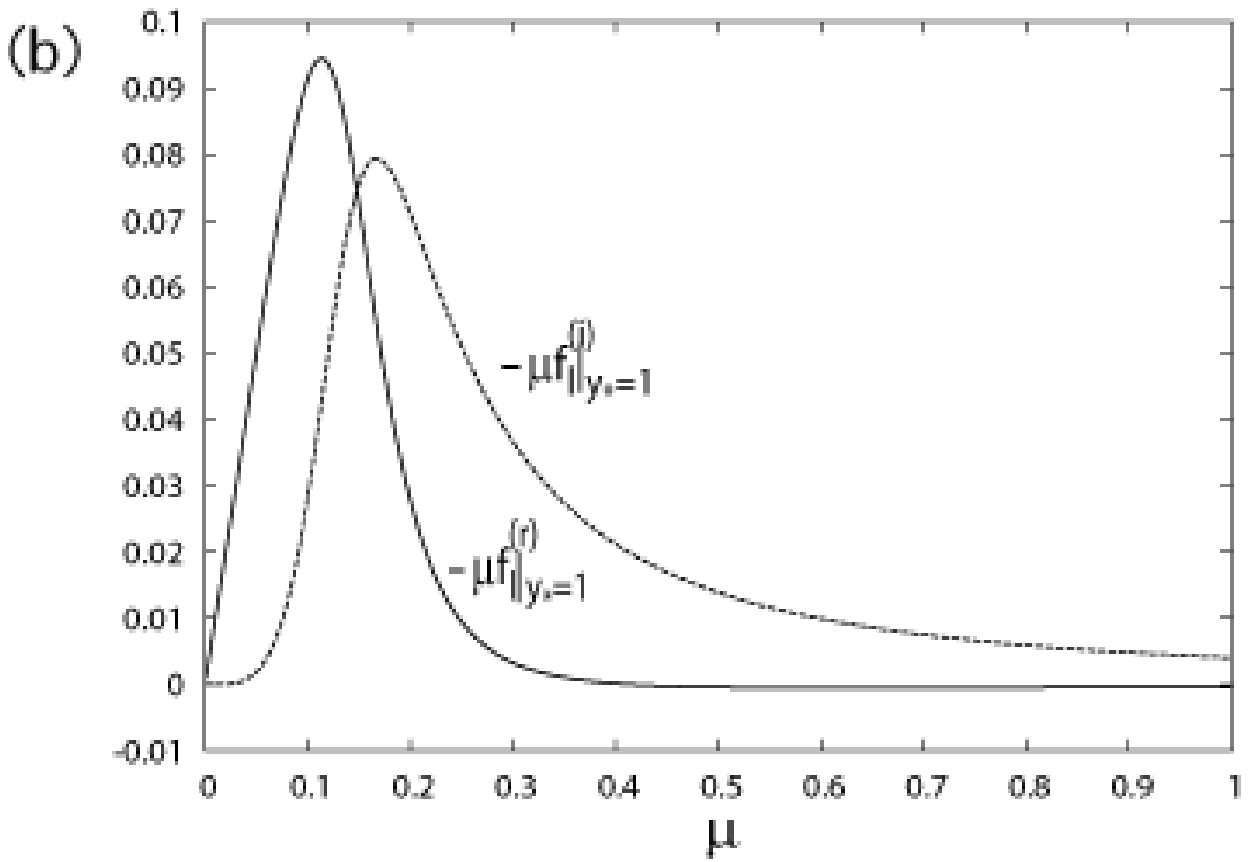}
\end{center}
\caption{For $Q/l=50$ [(ml/h)/cm] and $\theta=\pi/2$, (a) shows the dependence of the non-dimensional amplitude of the perturbed steam function in the water film, $|f_{l}|_{y_{*}=1}|$, and its real part $-f_{l}^{(r)}|_{y_{*}=1}$, imaginary part $-f_{l}^{(i)}|_{y_{*}=1}$, on the non-dimensional wave number $\mu$.  
(b) represents the behaviour of $-\mu f_{l}^{(r)}|_{y_{*}=1}$ and $-\mu f_{l}^{(i)}|_{y_{*}=1}$ against $\mu$. 
Here $\mu=1.0$ corresponds to the wavelength of 580 $\mu$m.}
\label{fig:famp-mu}
\end{figure}

A small perturbation of the ice-water interface in the non-dimensional form can be rewritten the following way: 
\begin{equation} 
y_{*}=\zeta_{*}=\delta_{b}\Imag[{\rm exp}(\sigma t+\rmi kx)]
=\delta_{b}(t)\sin[k(x-v_{p}t)], 
\label{eq:zeta}
\end{equation}
where $\delta_{b}=\zeta_{k}/h_{0}$, $\delta_{b}(t)\equiv \delta_{b}{\rm exp}(\sigma^{(r)}t)$, and $\Imag$ denotes the imaginary part of its argument. 
From the amplitude relation $\xi_{k}=-f_{l}|_{y_{*}=1}\zeta_{k}$, the corresponding perturbation of the water-air surface with an infinitesimal amplitude $\delta_{t}=\xi_{k}/h_{0}$ is given by
\begin{eqnarray} 
y_{*}=\xi_{*}&=&1+\Imag[\delta_{t}{\rm exp}(\sigma t+\rmi kx)] \nonumber \\
&=&1+|f_{l}|_{y_{*}=1}|\delta_{b}(t)\sin[k(x-v_{p}t)+\Theta_{\xi_{*}}],
\label{eq:xi}
\end{eqnarray}
where $|f_{l}|_{y_{*}=1}|=[(-f_{l}^{(r)}|_{y_{*}=1})^{2}+(-f_{l}^{(i)}|_{y_{*}=1})^2]^{1/2}$, 
$\cos\Theta_{\xi_{*}}=-f_{l}^{(r)}|_{y_{*}=1}/|f_{l}|_{y_{*}=1}|$ and
$\sin\Theta_{\xi_{*}}=-f_{l}^{(i)}|_{y_{*}=1}/|f_{l}|_{y_{*}=1}|$,
$\Theta_{\xi_{*}}$ being a phase difference between the ice-water interface and the water-air surface. When $\mu$ is small, $f_{l}|_{{y_{*}}=1}=f_{l}^{(r)}|_{y_{*}=1}+\rmi f_{l}^{(i)}|_{y_{*}=1}
\approx -36/(36+\alpha^{2})+\rmi (-6\alpha)/(36+\alpha^{2})$ (see equation (73) in (Ueno 2003)). 
When $f_{l}^{(i)}|_{{y_{*}}=1}$ acquires non-zero values, as shown in \fref{fig:famp-mu} (a), there exists a phase shift of the water-air surface against the ice-water interface.
The parameter $\alpha$ in (\ref{eq:alpha}) depends on $\mu$, hence the amplitude and phase of the water-air surface relative to the ice-water interface can change depending on the wave number of the ice-water interface (figures \ref{fig:famp-mu} (a) and \ref{fig:isotherm} (e)). 
Using the second equation in (\ref{eq:bc-Hl-Ueno}), the temperature gradient at the perturbed water-air surface can be expressed as 
$-\rmd T_{l}/\rmd y|_{y=\xi}=\bar{G}_{l}(1-\rmd H_{l}/\rmd y|_{y_{*}=1}\zeta)
=\bar{G}_{l}(1-\mu f_{l}|_{y_{*}=1}\zeta)$. 
\Fref{fig:famp-mu} (b) shows the behaviour of the real and imaginary part of $-\mu f_{l}|_{y_{*}=1}$ with respect to $\mu$. It is found that the perturbed part in $-\rmd T_{l}/\rmd y|_{y=\xi}$ increases with an increase in $\mu$, but that it is suppressed due to the restoring force acting on the water-air surface. This indicates that when the ice-water interface and water-air surface are coupled, even if the modes are sinusoidal in the linear stability analysis, the amplitude and phase shift of the water-air surface against the ice-water interface significantly influence the perturbed part of temperature in the water film through the boundary conditions in (\ref{eq:bc-Hl-Ueno}). 

On the other hand, the effect of the restoring force on the water-air surface was not taken into account in the model (Ogawa and Furukawa 2002), i.e., $f_{l}|_{y_{*}=1}=-1$, which means that the amplitude of the water-air surface is the same as that of the ice-water interface, and that there exists no phase shift of the water-air surface against the ice-water interface. Then, the perturbed part in $-\rmd T_{l}/\rmd y|_{y=\xi}=\bar{G}_{l}(1+\mu \zeta)$ only increases with $\mu$.
This is also the main difference between the two models.

\subsection{Dispersion relation \label{dis}}

Substituting $g_{l}(y)=H_{l}(y)\bar{G}_{l}\zeta_{k}$ and (\ref{eq:sol-gs-Ueno}) into the second equation in (\ref{eq:bco(1)-sl-T})
and using $K_{l}\bar{G}_{l}=K_{a}\bar{G}_{a}$ yields the dispersion relation for the perturbation of the ice-water interface:
\begin{equation}
\sigma=\frac{K_{a}\bar{G}_{a}}{Lh_{0}}\left\{-\frac{\rmd H_{l}}{\rmd y_{*}}\Big|_{y_{*}=0}
+K^{s}_{l}\mu (-G^{s}_{l}+H_{l}|_{y_{*}=0}-1)\right\},
\label{eq:dispersion}
\end{equation}
where $K_{l}^{s}\equiv K_{s}/K_{l}=3.96$ is the ratio of the thermal conductivity of ice to that of water and 
$G^{s}_{l}\equiv \bar{G}_{s}/\bar{G}_{l}=(K_{l}/K_{a})(\delta_{0}/b_{0})(T_{sl}-T_{\rm sub})/(T_{sl}-T_{\infty})$ is the ratio of the unperturbed temperature gradient at the ice-water interface in the ice to that in the water. The real and imaginary part of (\ref{eq:dispersion}) yield the non-dimensional amplification rate $\sigma_{*}^{(r)}\equiv \sigma^{(r)}/\{K_{a}\bar{G}_{a}/(Lh_{0})\}$ and the non-dimensional phase velocity $v_{p*}=v_{p}/(K_{a}\bar{G}_{a}/L)=-\sigma^{(i)}/(kK_{a}\bar{G}_{a}/L)$, respectively, 
\begin{equation}
\sigma_{*}^{(r)}=-\frac{\rmd H_{l}^{(r)}}{\rmd y_{*}}\Big|_{y_{*}=0}+K^{s}_{l}\mu\left(-G^{s}_{l}+H_{l}^{(r)}|_{y_{*}=0}-1\right),
\label{eq:amp}
\end{equation}
\begin{equation}
v_{p*}=-\frac{1}{\mu}\left(-\frac{\rmd H_{l}^{(i)}}{\rmd y_{*}}\Big|_{y_{*}=0}+K^{s}_{l}\mu H_{l}^{(i)}|_{y_{*}=0}\right),
\label{eq:vp}
\end{equation}
where $H_{l}^{(r)}$ and $H_{l}^{(i)}$ are the real and imaginary part of $H_{l}$.

Equations (\ref{eq:gov-fl}) and (\ref{eq:gov-Hl}) were solved analytically with the boundary conditions (\ref{eq:bc-fl}) and (\ref{eq:bc-Hl-Ueno}) under the long wavelength approximation neglecting the higher order of $\mu$, except for retaining the second term in $\alpha$ because of $(a/h_{0})^{2} \gg 1$ (Ueno 2003).
By transferring the variable $y_{*}$ into $z=1-y_{*}$, the general solution of (\ref{eq:gov-Hl}) can be expressed as 
$H_{l}(z)
=C_{1}\phi_{1}(z)+C_{2}\phi_{2}(z) 
+\rmi \mu \Pec_{l}
\int_{0}^{z}\left\{\phi_{2}(z)\phi_{1}(z')-\phi_{1}(z)\phi_{2}(z')\right\}f_{l}(z')\rmd z'$, 
where $C_{1}$ and $C_{2}$ are unknown constants, $\phi_{1}(z)$ and $\phi_{2}(z)$ are the homogeneous solutions of (\ref{eq:gov-Hl}): $\rmd^{2}\phi/\rmd z^{2}-\{\mu^{2}+\rmi \mu\Pec_{l}\bar{U}_{l*}(z)\}\phi=0$, where $\bar{U}_{l*}(z)=1-z^{2}$. 
We note that in the $z$ coordinate, $z=0$ and $z=1$ are positions of the unperturbed water-air surface and ice-water interface, respectively. 
Neglecting the $\mu^{2}$ term in the above homogeneous equation, $\phi_{1}(z)$ and $\phi_{2}(z)$ can be expanded in terms of $\mu\Pec_{l}$ as follows (see APPENDIX in (Ueno 2003)):
\begin{equation}
\fl
\phi_{1}(z) 
=1+\rmi\left(\frac{1}{2}z^{2}-\frac{1}{12}z^{4}\right)\mu\Pec_{l} 
+\left(-\frac{1}{24}z^{4}+\frac{7}{360}z^{6}-\frac{1}{672}z^{8}\right)(\mu \Pec_{l})^2
+\cdots, 
\label{eq:phi1}
\end{equation}
\begin{equation}
\fl
\phi_{2}(z) 
=z+\rmi\left(\frac{1}{6}z^{3}-\frac{1}{20}z^{5}\right)\mu \Pec_{l} 
+\left(-\frac{1}{120}z^{5}+\frac{13}{2520}z^{7}-\frac{1}{1440}z^{9}\right)(\mu \Pec_{l})^{2}
+\cdots .
\label{eq:phi2}
\end{equation}

The boundary conditions in (\ref{eq:bc-Hl-Ueno}) give $C_{1}=-f_{l}|_{z=0}$ and $C_{2}=-\mu f_{l}|_{z=0}$, respectively, because $\phi_{1}|_{z=0}=1$, $\phi_{2}|_{z=0}=0$, $\rmd\phi_{1}/\rmd z|_{z=0}=0$ and $\rmd\phi_{2}/\rmd z|_{z=0}=1$. Thus we obtain 
\begin{equation}
\fl
H_{l}(z)
=-f_{l}|_{z=0}\left\{\phi_{1}(z)+\mu\phi_{2}(z)\right\} 
+\rmi \mu \Pec_{l}
\int_{0}^{z}\left\{\phi_{2}(z)\phi_{1}(z')-\phi_{1}(z)\phi_{2}(z')\right\}f_{l}(z')\rmd z'.
\label{eq:generalsol-Hl}
\end{equation}
Since
\begin{eqnarray}
\phi_{1}|_{z=1}=1+\rmi\frac{5}{12}(\mu\Pec_{l})-\frac{239}{10080}(\mu\Pec_{l})^{2}+\cdots, \nonumber \\
\phi_{2}|_{z=1}=1+\rmi\frac{7}{60}(\mu\Pec_{l})-\frac{13}{3360}(\mu\Pec_{l})^{2}+\cdots, \nonumber \\
\frac{\rmd\phi_{1}}{\rmd z}\Big|_{z=1}=\rmi\frac{2}{3}(\mu\Pec_{l})-\frac{13}{210}(\mu\Pec_{l})^{2}+\cdots, \nonumber \\
\frac{\rmd\phi_{2}}{\rmd z}\Big|_{z=1}=1+\rmi\frac{1}{4}(\mu\Pec_{l})-\frac{17}{1440}(\mu\Pec_{l})^{2}+\cdots, 
\label{eq:phi12}
\end{eqnarray}
the ratios of the second order term in $\mu\Pec_{l}$ to the first order one in $\phi_{1}|_{z=1}$, $\phi_{2}|_{z=1}$, $\rmd\phi_{1}/\rmd z|_{z=1}$ and $\rmd\phi_{2}/\rmd z|_{z=1}$ are about $5.7\times 10^{-2}\mu\Pec_{l}$, $3.3\times 10^{-2}\mu\Pec_{l}$, $9.3\times 10^{-2}\mu\Pec_{l}$ and $4.7\times 10^{-2}\mu\Pec_{l}$, respectively. The second order terms in $\mu\Pec_{l}$ are not negligible when $\mu\Pec_{l} \sim 10$. This is possible when $Q/l \sim 300$ [(ml/h)/cm] for the wavelength of ripples on icicles. However, the typical values of $Q$ over icicles are on the order of tens of ml/h and their radii are usually in the range of $1\sim 10$ cm, the value of $Q/l$ is in the range $10 \sim 100$ [(ml/h)/cm] (Maeno \etal 1994, Short \etal 2006). 
As far as we are limited to such a range of $Q/l$, $\mu\Rey_{l} \ll 1$ and $\mu\Pec_{l}\sim 1$ for the length scale of ripples on icicles. We can neglect the $\mu\Rey_{l}$ term in (\ref{eq:gov-fl}) and (\ref{eq:bc-fl}). This corresponds to neglecting the inertia term of the full Orr-Sommerfeld equation, then we can approximate (\ref{eq:gov-fl}) as follows: $\rmd^{4}f_{l}/\rmd y_{*}^{4}=0$. The solution of this equation with the boundary conditions in (\ref{eq:bc-fl}) is given by
$f_{l}(z)=(-6+\rmi \alpha z+6z^{2}-\rmi \alpha z^{3})/(6-\rmi \alpha)$ (Ueno 2003, Ueno 2007).
As far as $\mu\Pec_{l}\sim 1$, it is sufficient to consider up to the first order in $\mu\Pec_{l}$ in (\ref{eq:phi1}) and (\ref{eq:phi2}) because the second order terms in $\mu\Pec_{l}$ in (\ref{eq:phi12}) are very small compared to the first order terms in $\mu\Pec_{l}$ as estimated above. 

Furthermore, if the substrate is not sufficiently cold or if there is not heat conduction through the substrate, the unperturbed part of heat conduction to the interior of the icicle is negligible (Makkonen 1988). The semi-infinite ice layer approximation of previous paper (Ueno 2003) corresponds to the case of $b_{0} \gg 1$. In these situations, $G^{s}_{l}=0$,  hence (\ref{eq:amp}) and (\ref{eq:vp}) yield (Ueno 2003)
\begin{eqnarray}
\fl
\sigma_{*}^{(r)}
= 
\frac{\mu\left\{36-\frac{3}{2}\alpha(\mu\Pec_{l})\right\}-\frac{3}{2}\alpha(\mu\Pec_{l})}{36+\alpha^{2}} 
+K^{s}_{l}\mu
\frac{\mu\left\{36-\frac{7}{10}\alpha(\mu\Pec_{l})\right\}-\frac{7}{10}\alpha(\mu\Pec_{l})-\alpha^{2}}{36+\alpha^{2}},\nonumber \\
\label{eq:amp-Ueno}
\end{eqnarray}
\begin{eqnarray}
\fl
v_{p*}=-\frac{1}{\mu}
\left[\frac{\mu\left\{6\alpha+9(\mu\Pec_{l})\right\}-\frac{1}{4}\alpha^{2}(\mu\Pec_{l})}{36+\alpha^{2}} \right. \nonumber \\
\left.
+K^{s}_{l}\mu\frac{\mu\left\{6\alpha+\frac{21}{5}(\mu\Pec_{l})\right\}+6\alpha-\frac{7}{60}\alpha^{2}(\mu\Pec_{l})}{36+\alpha^{2}}\right].
\label{eq:vp-Ueno}
\end{eqnarray}
It should be noted that in the case of $G^{s}_{l}=0$, $\sigma^{(r)}_{*}$ and $v_{p*}$ are independent of the unperturbed air temperature gradient $\bar{G}_{a}$.

On the other hand, $H_{l}|_{y_{*}=0}=H_{l}|_{z=1}=1$ in the case of $\Delta T_{sl}=0$ and $f_{l}|_{y_{*}=1}=-1$ (Ogawa and Furukawa 2002) and noting that $K_{a}/K_{l} \ll 1$, the solution $H_{l}$ with the boundary conditions in (\ref{eq:bc-Hl-OF}) is given by  
$H_{l}(z)
=\{(1-\rmi \mu\Pec_{l} I|_{z=1})\phi_{1}(z)+\mu(\phi_{1}|_{z=1}\phi_{2}(z)-\phi_{2}|_{z=1}\phi_{1}(z))\}/\phi_{1}|_{z=1}+\rmi \mu\Pec_{l} I(z)$,
where $I(z)=\int_{0}^{z}\{\phi_{2}(z)\phi_{1}(z')-\phi_{1}(z)\phi_{2}(z')\}f_{l}(z')\rmd z'$ (Ueno 2004).
This solution gives $-\rmd H_{l}/\rmd y_{*}|_{y_{*}=0}=\rmd H_{l}/\rmd z|_{z=1}=\mu/\phi_{1}|_{z=1}$ and in the case of $G^{s}_{l}=0$, (\ref{eq:amp}) and (\ref{eq:vp}) yield
\begin{eqnarray}
\sigma_{*}^{(r)} =\frac{\mu\{1-\frac{239}{10080}(\mu\Pec_{l})^2\}}{\left\{1-\frac{239}{10080}(\mu\Pec_{l})^{2}\right\}^{2}+\left\{\frac{5}{12}\mu\Pec_{l}\right\}^{2}},
\label{eq:amp-OF}
\end{eqnarray}
\begin{eqnarray}
v_{p*} 
=\frac{\frac{5}{12}\mu\Pec_{l}}{\left\{1-\frac{239}{10080}(\mu\Pec_{l})^{2}\right\}^{2}+\left\{\frac{5}{12}\mu\Pec_{l}\right\}^{2}}.
\label{eq:vp-OF}
\end{eqnarray}
We notice that (\ref{eq:amp-OF}) and (\ref{eq:vp-OF}) are the same results as equations (77) and (81) in (Ogawa and Furukawa 2002), except $\mu\Pec_{l}$ is defined as $\alpha$ in their paper. It is remarked that (\ref{eq:amp-OF}) and (\ref{eq:vp-OF}) are obtained by expanding $\phi_{1}|_{z=1}$ up to the second order in $\mu\Pec_{l}$, in contrast to (\ref{eq:amp-Ueno}) and (\ref{eq:vp-Ueno}). 

\section{Comparison of analytical with numerical results \label{comp}}

In spite of using many approximations in previous papers (Ogawa and Furukawa 2002, Ueno 2003), the analytical calculations to solve the equations for $f_{l}$ and $H_{l}$ with appropriate boundary conditions were very complex and cumbersome. 
Instead, by decomposing $f_{l}$ and $H_{l}$ into its real part $f_{l}^{(r)}, H_{l}^{(r)}$, imaginary part $f_{l}^{(i)}, H_{l}^{(i)}$, we performed numerical studies to solve the ordinary differential equations for $f_{l}^{(r)}$, $f_{l}^{(i)}$, $H_{l}^{(r)}$ and $H_{l}^{(i)}$ in equations (\ref{eq:gov-fl}) and (\ref{eq:gov-Hl}), with boundary conditions (\ref{eq:bc-fl}) and (\ref{eq:bc-Hl-Ueno}) and without approximations as used to derive the analytical results mentioned above. 

\begin{figure}[t]
\begin{center}
\includegraphics[width=7cm,height=7cm,keepaspectratio,clip]{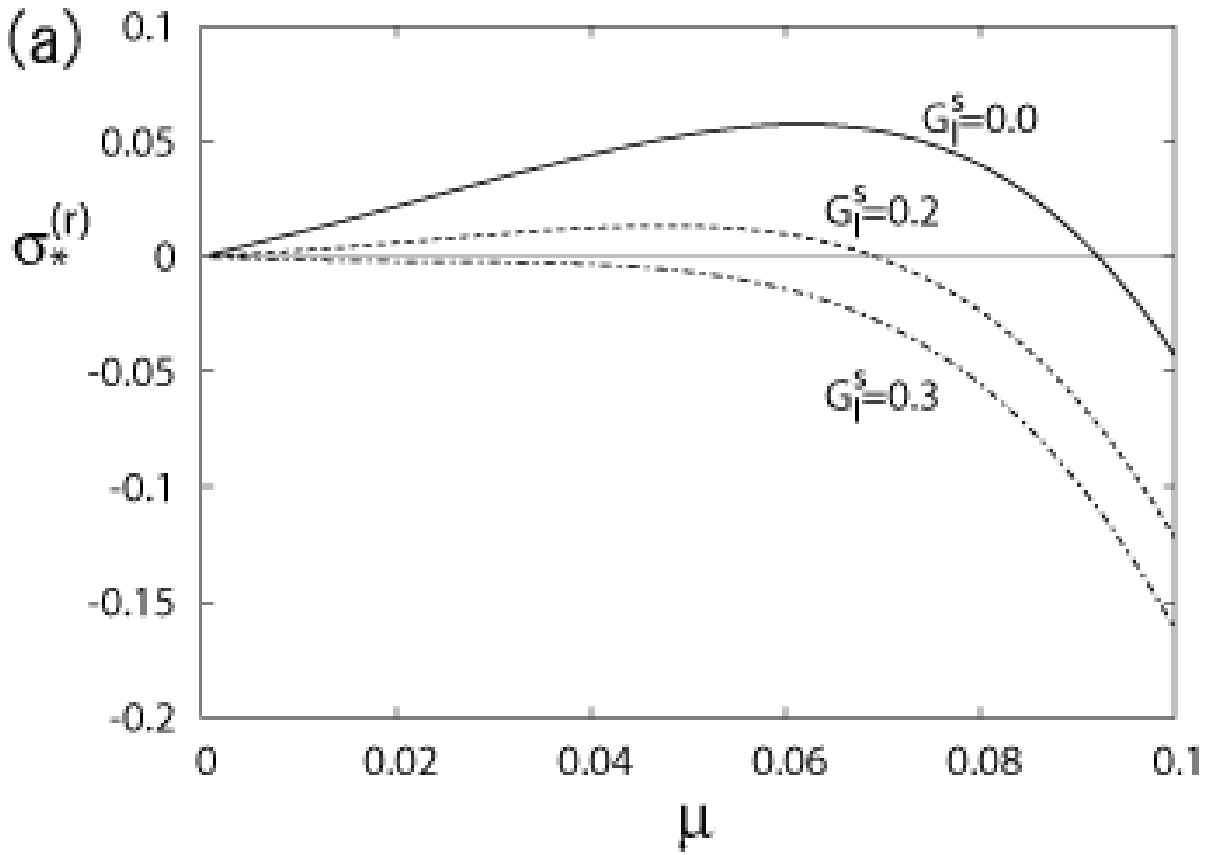}\\[5mm]
\includegraphics[width=7cm,height=7cm,keepaspectratio,clip]{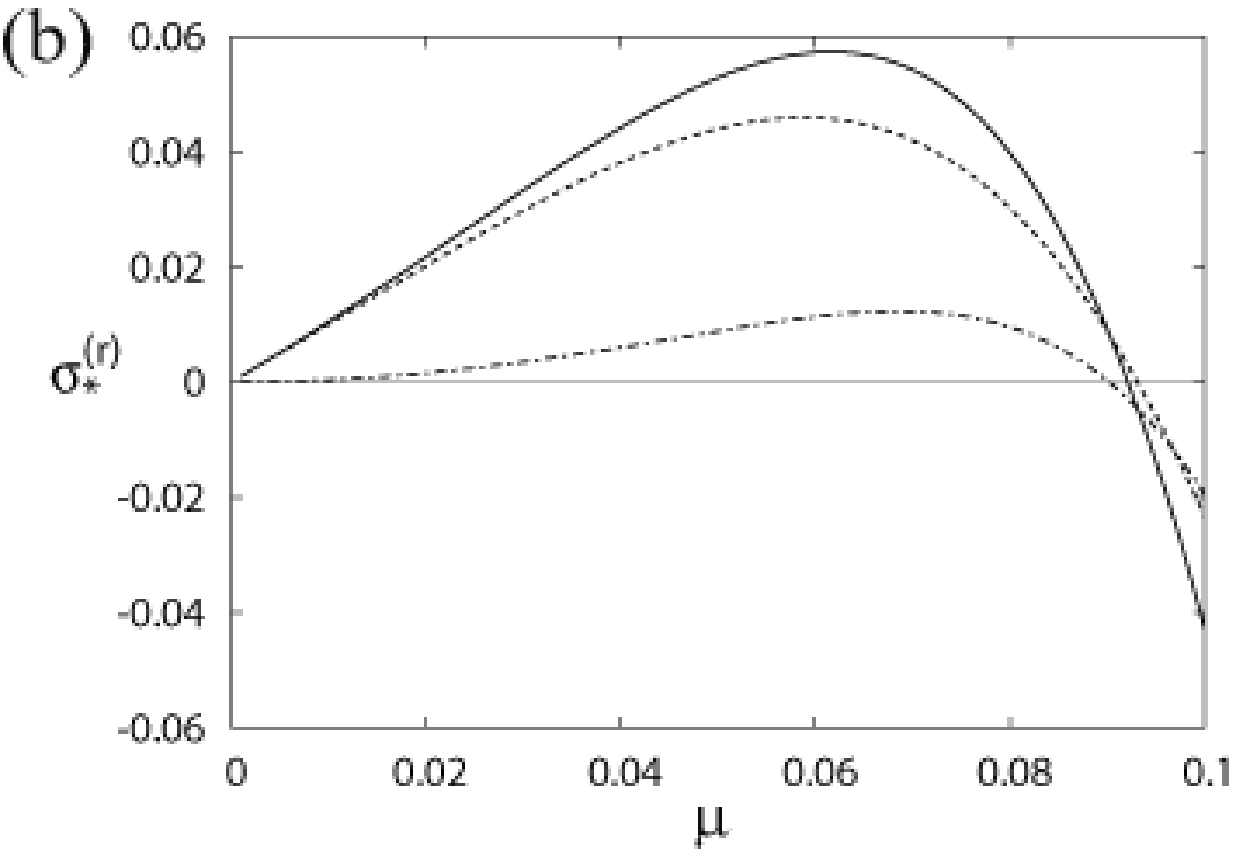}\hspace{5mm}
\includegraphics[width=7cm,height=7cm,keepaspectratio,clip]{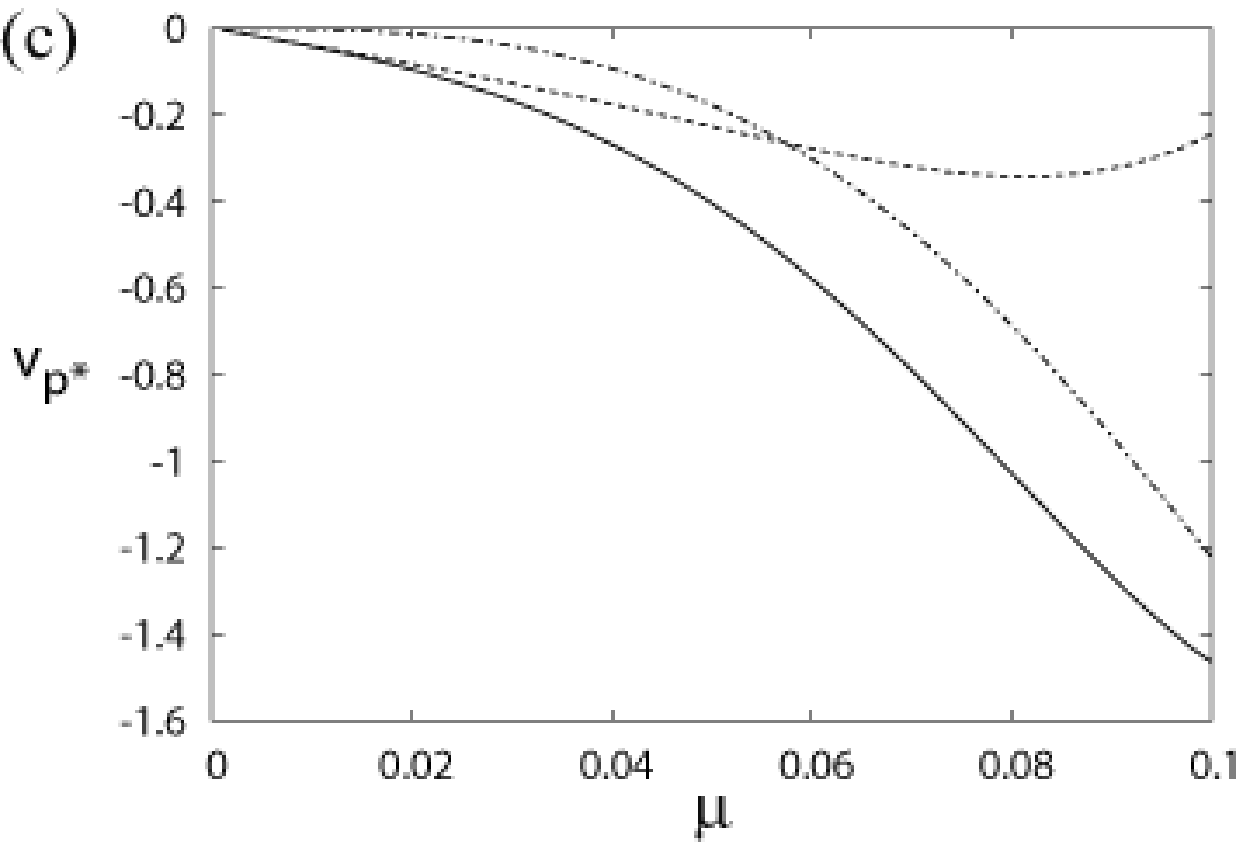}
\end{center}
\caption{Non-dimensional amplification rate $\sigma_{*}^{(r)}$ versus 
non-dimensional wave number $\mu$ at $Q/l=50$ [(ml/h)/cm] and $\theta=\pi/2$.
(a) is for different values of $G^{s}_{l}$. 
(b) and (c) are in the case of $G^{s}_{l}=0$.
Solid lines: $\sigma^{(r)}_{*}$ and $v_{p*}$ are obtained from (\ref{eq:amp}) and (\ref{eq:vp}), respectively;
dashed lines: the contribution of the first term in (\ref{eq:amp}) and (\ref{eq:vp});
dashed-dotted lines: the contribution of the second term in (\ref{eq:amp}) and (\ref{eq:vp}).
Here $\mu=0.1$ corresponds to the wavelength of 5.8 mm.}
\label{fig:ampGsl-ampc-vpc}
\end{figure}

First, in the case of $G^{s}_{l}\neq 0$, we have to consider heat conduction through a finite ice thickness into the substrate.
\Fref{fig:ampGsl-ampc-vpc} (a) shows the non-dimensional amplification rate $\sigma_{*}^{(r)}$ versus the non-dimensional wave number $\mu$ for different values of $G^{s}_{l}$ at $Q/l=50$ [(ml/h)/cm] and $\theta=\pi/2$. In the range of $0<G^{s}_{l}<0.3$, the wavelengths are longer than that in the case of $G^{s}_{l}=0$ as the value of $G^{s}_{l}$ increases. We find that $\sigma_{*}^{(r)}<0$ for all $\mu$ above $G^{s}_{l}=0.3$. This means that if we choose the parameters in $G^{s}_{l}=(K_{l}/K_{a})(\delta_{0}/b_{0})(T_{sl}-T_{\rm sub})/(T_{sl}-T_{\infty})$ to satisfy $G^{s}_{l}>0.3$, ripples do not appear on the ice surface. 
Indeed, this is relevant as an experimental result that no ripples were observed on the ice grown on an planar aluminum substrate by supplying water from the top of the apparatus in a cold room at temperature below 0 $^{\circ}$C (Matsuda 1997). A smooth ice surface was produced on the aluminum substrate. Matsuda states that since the thermal conductivity of aluminum is about 100 times greater than that of ice, most of the latent heat released at the ice-water interface is conducted into the aluminum substrate through the ice. If there exists heat conduction through the substrate, the continuity of heat flux at the boundary between substrate and ice is $K_{\rm sub}(T_{\rm sub}-T_{\rm sub0})/l_{\rm sub}=K_{s}(T_{sl}-T_{\rm sub})/b_{0}$, where $K_{\rm sub}$ is the thermal conductivity of a substrate, $l_{\rm sub}$ is the thickness of the substrate, $T_{\rm sub}$ and $T_{\rm sub0}$ are temperatures at the boundary between substrate and ice and that at the other side of the substrate, respectively. From this, 
$T_{\rm sub}=\{T_{\rm sub0}+(K_{s}/K_{\rm sub})(l_{\rm sub}/b_{0})T_{sl}\}/\{1+(K_{s}/K_{\rm sub})(l_{\rm sub}/b_{0})\}$ is obtained. Substituting this $T_{\rm sub}$ into the above $G^{s}_{l}$, we obtain
$G^{s}_{l}=(K_{l}/K_{a})(\delta_{0}/b_{0})(T_{sl}-T_{\rm sub0})/(T_{sl}-T_{\infty})1/\{1+(K_{s}/K_{\rm sub})(l_{\rm sub}/b_{0})\}$.
When the thickness of ice grown on the planar aluminum substrate satisfies the condition $b_{0} \gg (K_{s}/K_{\rm sub})l_{\rm sub}$, the above $G^{s}_{l}$ can be approximated as
$G^{s}_{l}=(K_{l}/K_{a})(\delta_{0}/b_{0})(T_{sl}-T_{\rm sub0})/(T_{sl}-T_{\infty})$. 
Moreover if the other side of surface of the aluminum substrate is exposed to ambient cold air, we assume $T_{\rm sub0}=T_{\infty}$. Then $G^{s}_{l}=(K_{l}/K_{a})(\delta_{0}/b_{0})$ satisfies the condition $G^{s}_{l}>0.3$ when $b_{0}<100$ $\delta_{0}$.
While the thickness of ice growing on the planar aluminum substrate by supplying water satisfies this condition, ripples would not appear on the ice surface.

In the following discussions and the next sections, we will focus on the case of $G^{s}_{l}=0$. 
The solid lines in figures \ref{fig:ampGsl-ampc-vpc} (b) and (c) show
$\sigma_{*}^{(r)}$ in (\ref{eq:amp}) and $v_{p*}$ in (\ref{eq:vp}) against $\mu$ at $Q/l=50$ [(ml/h)/cm] and $\theta=\pi/2$, respectively. 
The dashed and dashed-dotted lines in figures \ref{fig:ampGsl-ampc-vpc} (b) and (c) represent the first and second term in (\ref{eq:amp}) and (\ref{eq:vp}), respectively. 
In the case of $\Delta T_{sl}=0$ (Ogawa and Furukawa 2002), the second terms in (\ref{eq:amp}) and (\ref{eq:vp}) must be equal to zero because $H_{l}^{(r)}|_{y_{*}=0}=1$ and $H_{l}^{(i)}|_{y_{*}=0}=0$ from the first equation in (\ref{eq:bc-Hl-OF}). On the other hand, the contribution of $\Delta T_{sl} \neq 0$ in the model (Ueno 2003) appears in the second terms in (\ref{eq:amp-Ueno}) and (\ref{eq:vp-Ueno}). Indeed, as shown by the dashed-dotted line in figures \ref{fig:ampGsl-ampc-vpc} (b) and (c), the contribution of the second terms to the total values of $\sigma_{*}^{(r)}$ and $v_{p*}$ is not negligible. However, we can make an approximation in the case of figure \ref{fig:ampGsl-ampc-vpc} (b) because the second term (dashed-dotted line) is smaller than the first term (dashed line) and the wave number at which $\sigma_{*}^{(r)}$ acquires a maximum value is almost the same as that without the second term. 
When $\mu$ is small, we can approximate (\ref{eq:amp}) as follows: 
$\sigma_{*}^{(r)} 
\approx \rmd H_{l}^{(r)}/\rmd z|_{z=1}
\approx -\mu f_{l}^{(r)}|_{z=0}
+(2/3)\mu\Pec_{l}f_{l}^{(i)}|_{z=0}
-\mu\Pec_{l}\int_{0}^{1}f_{l}^{(i)}(z)\rmd z$. 
This indicates that the perturbed part of temperature gradient at the ice-water interface is affected by the amplitude $f_{l}^{(r)}|_{z=0}$ and $f_{l}^{(i)}|_{z=0}$ of the perturbed part of stream function at the water-air surface. 
Since $\mu\Pec_{l} \sim 1$ and $\alpha \sim 1$ for small $\mu$ and the typical range of $Q/l=10 \sim 100$ [(ml/h)/cm] and  $\theta=\pi/2$, extracting the most dominant term from $\rmd H_{l}^{(r)}/\rmd z|_{z=1}$ and using (\ref{eq:alpha}), we obtain 
\begin{equation}
\sigma_{*}^{(r)}
\approx 
\frac{36\mu-\frac{3}{2}\alpha(\mu\Pec_{l})}{36}=\mu-\frac{\Pec_{l}}{12}\left(\frac{a}{h_{0}}\right)^{2}\mu^{4}.
\label{eq:amp-Ueno-approx}
\end{equation}

A positive destabilizing term in (\ref{eq:amp-Ueno-approx}) is derived from 
the first term in $\rmd H_{l}^{(r)}/\rmd z|_{z=1}$.
We find that when $\mu$ is small, the trigger of the destabilization of the ice-water interface originates from 
the perturbed part of air temperature gradient at the water-air surface
because $-\mu f_{l}^{(r)}|_{z=0}$ of the second equation in (\ref{eq:bc-Hl-Ueno}) is proportional to $\mu$ when $\mu$ is small as shown in \fref{fig:famp-mu} (b). 
On the other hand, a negative stabilizing term in (\ref{eq:amp-Ueno-approx}) is derived from the sum of second and third terms in $\rmd H_{l}^{(r)}/\rmd z|_{z=1}$.
As shown in \fref{fig:famp-mu} (a), as $\mu$ increases, $f_{l}^{(i)}$ acquires non-zero values, hence the sum of second and third terms in $\rmd H_{l}^{(r)}/\rmd z|_{z=1}$ dominates over the first term and suppresses the instability. When $f_{l}^{(i)}|_{{y_{*}}=1} \neq 0$, we showed that there exists a phase shift of the water-air surface against the ice-water interface. This suggests that the instability and/or stability of disturbances of the ice-water interface is related to the magnitude of phase shift of the water-air surface, which will be discussed in the next section.  
As a result of competition between the first and second term in (\ref{eq:amp-Ueno-approx}), we find from $\rmd\sigma_{*}^{(r)}/\rmd\mu=0$ that $\sigma_{*}^{(r)}$ acquires a maximum value at $\mu=[3(h_{0}/a)^2/\Pec_{l}]^{1/3}$. From this, we obtain a simpler formula to determine the wavelength of ripples: $\lambda=2\pi(a^{2}h_{0}\Pec_{l}/3)^{1/3}$ (for the dependence of $\lambda$ on $Q/l$, see Fig. 6 (a) in (Ueno 2007)). 
This formula includes two characteristic lengths $a$ and $h_{0}$. Indeed, using the typical values of $a=2.8$ mm, $h_{0} \sim 100$ $\mu$m and $\Pec_{l} \sim 10$, one centimeter scale wavelength is obtained from the above formula. It should be noted that this long-length scale is in contrast with the wavelength $\lambda_{\rm MS}=2\pi\sqrt{l_{d}d_{0}}$ obtained from the Mullins-Sekerka theory, which is of order microns (Mullins-Sekerka 1963). Here, $l_{d}$ is the thermal diffusion length and is usually a macroscopic length, whereas $d_{0}$ is the capillary length associated with the solid-liquid interface tension, and is a microscopic length of order angstroms (Langer 1980, Caroli \etal 1992).    

\begin{figure}[t]
\begin{center}
\includegraphics[width=7cm,height=7cm,keepaspectratio,clip]{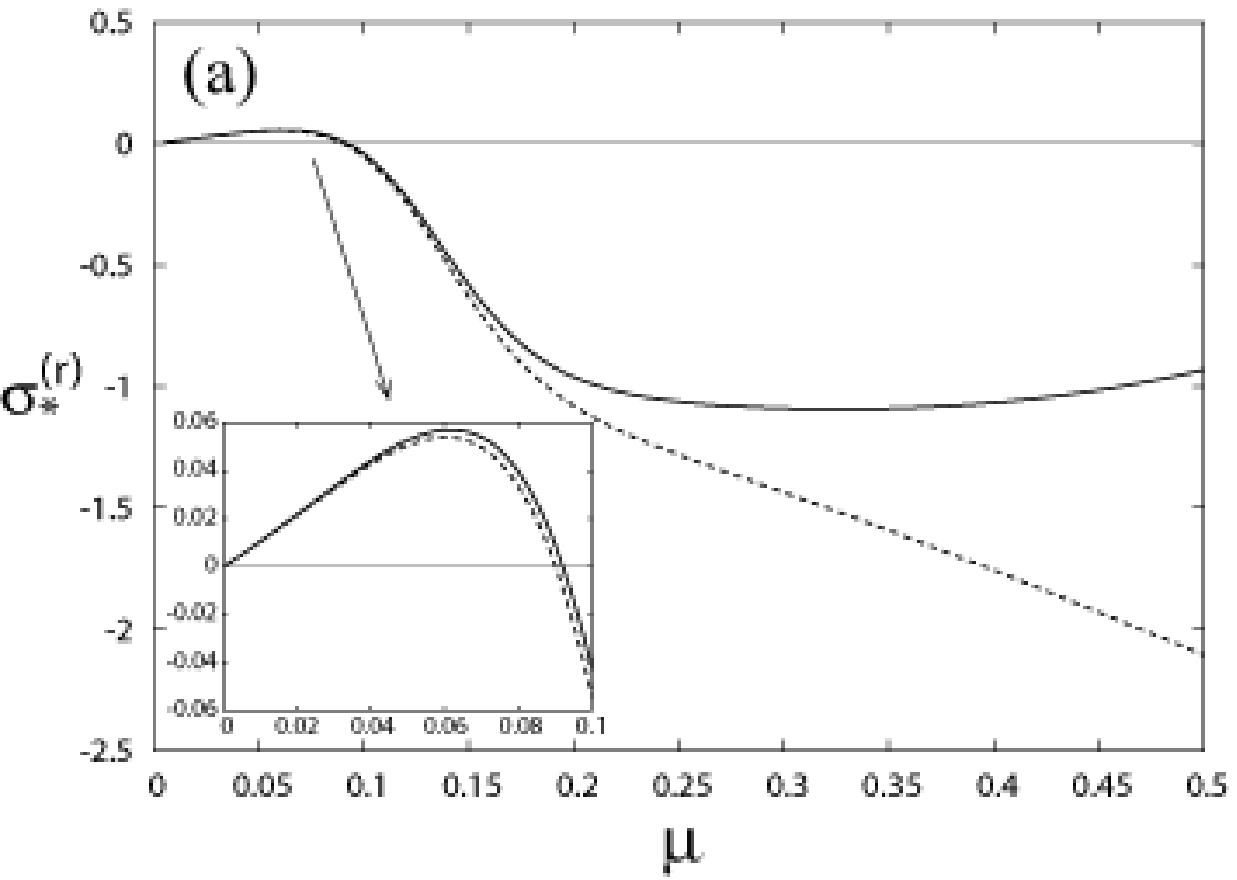}\hspace{5mm}
\includegraphics[width=7cm,height=7cm,keepaspectratio,clip]{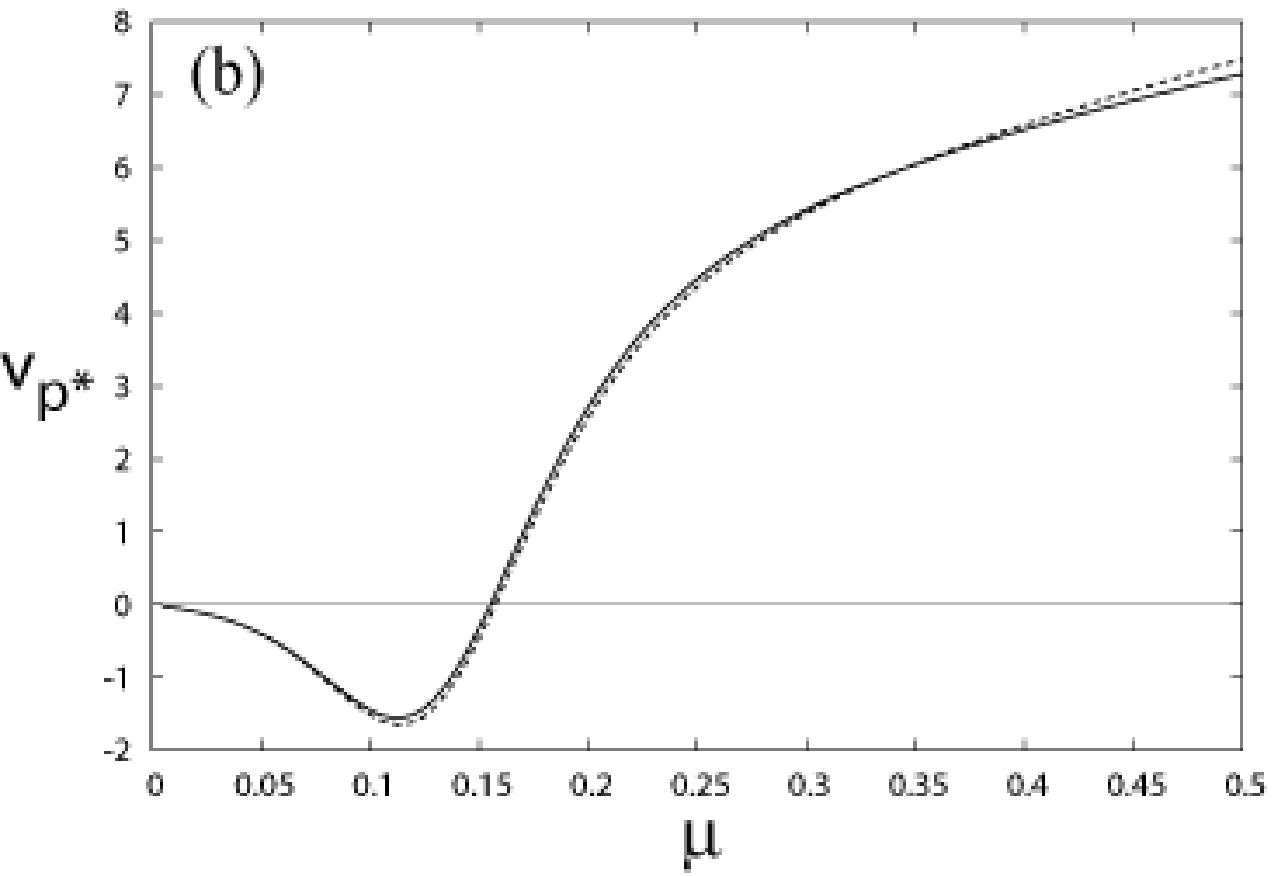}\\[0.5cm]
\includegraphics[width=7cm,height=7cm,keepaspectratio,clip]{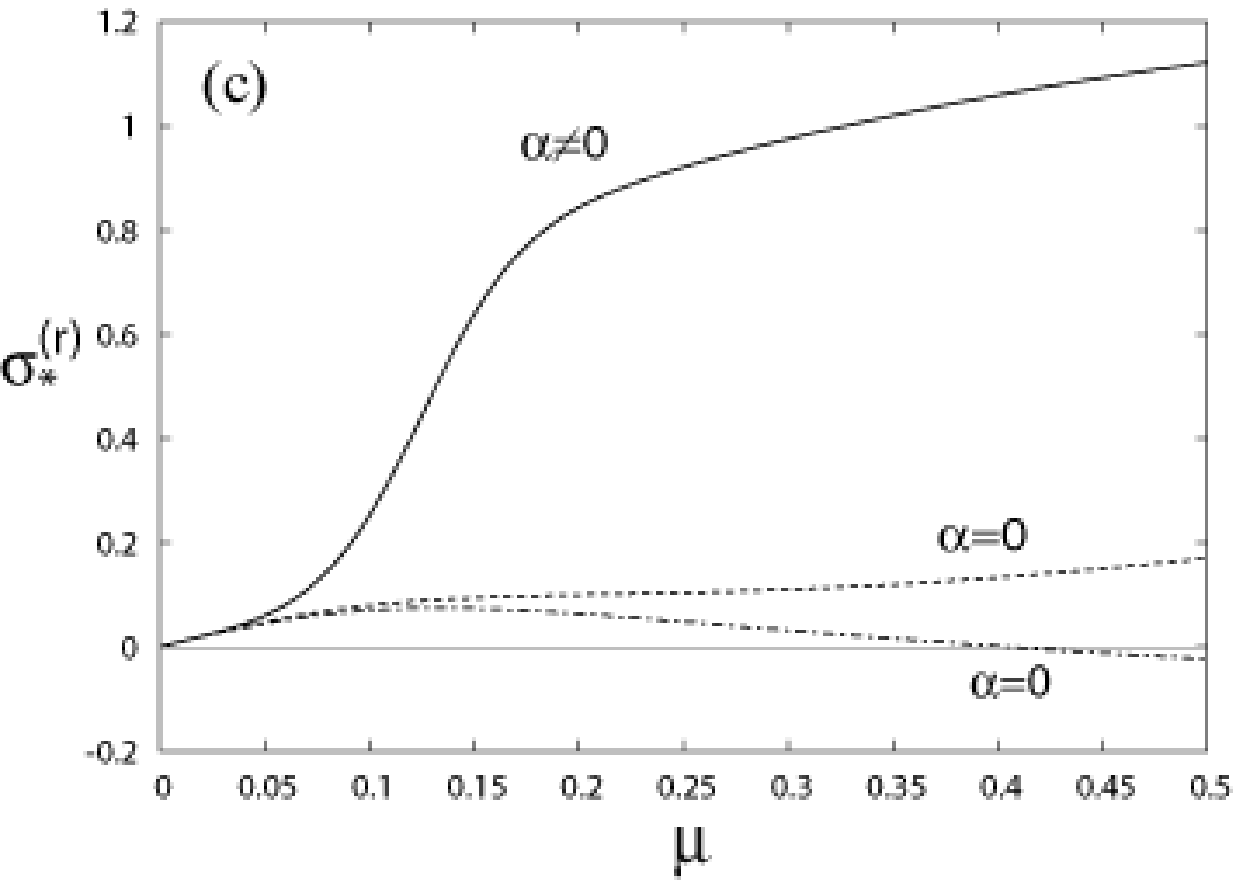}\hspace{5mm}
\includegraphics[width=7cm,height=7cm,keepaspectratio,clip]{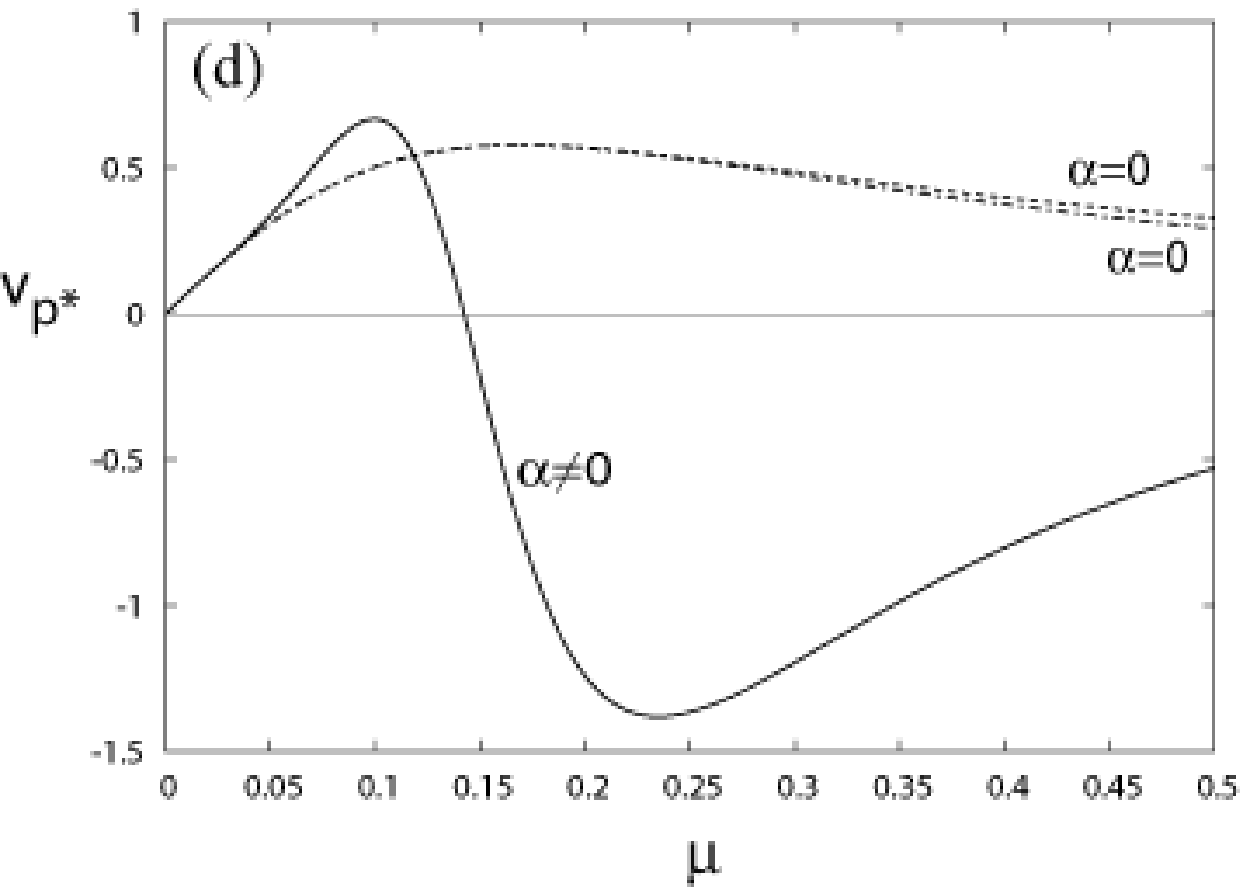}
\end{center}
\caption{Non-dimensional amplification rate $\sigma_{*}^{(r)}$ and non-dimensional phase velocity $v_{p*}$ versus 
non-dimensional wave number $\mu$ at $Q/l=50$ [(ml/h)/cm] and $\theta=\pi/2$. 
(a) and (b) are in the case of $\Delta T_{sl} \neq 0$ and $\Delta T_{la}=0$. 
(c) and (d) are in the case of $\Delta T_{sl}=0$ and $\Delta T_{la} \neq 0$.
Solid lines in (a) and (b): numerical results with boundary conditions in (Ueno 2003); 
dashed lines: analytical results (\ref{eq:amp-Ueno}) and (\ref{eq:vp-Ueno}). 
Solid lines in (c) and (d) $(\alpha \neq 0)$ and dashed lines $(\alpha=0)$: numerical results calculated by us with boundary conditions in (Ogawa and Furukawa 2002); 
dashed-dotted lines $(\alpha=0)$: analytical results (\ref{eq:amp-OF}) and (\ref{eq:vp-OF}).
Here $\mu=0.5$ corresponds to the wavelength of 1.2 mm.}
\label{fig:theory-numerical}
\end{figure}

In the case of $T_{s}|_{y=\zeta}=T_{l}|_{y=\zeta}=T_{sl}+\Delta T_{sl}$ and $T_{l}|_{y=\xi}=T_{a}|_{y=\xi}=T_{la}$ (Ueno 2003),  
the numerical results are shown by the solid lines in figures \ref{fig:theory-numerical} (a) and (b). The dashed lines are the analytical results, (\ref{eq:amp-Ueno}) and (\ref{eq:vp-Ueno}). The deviation of the dashed line from the solid line in \fref{fig:theory-numerical} (a) is mainly due to the neglect of the higher order of $\mu\Pec_{l}$. However, as shown in the inset in \fref{fig:theory-numerical} (a), the analytical result is in good agreement with the numerical result as far as we are concerned with the long wavelength region of $\mu<0.15$. Here $\mu=0.15$ corresponds to the wavelength of 3.8 mm for $Q/l=50$ [(ml/h)/cm] and $\theta=\pi/2$. 

In the case of $T_{s}|_{y=\zeta}=T_{l}|_{y=\zeta}=T_{sl}$ and $T_{l}|_{y=\xi}=T_{a}|_{y=\xi}=T_{la}+\Delta T_{la}$ (Ogawa and Furukawa 2002), the same ordinary differential equations were solved numerically with the same boundary conditions by replacing only the first equation in (\ref{eq:bc-Hl-Ueno}) with that in (\ref{eq:bc-Hl-OF}), and by neglecting the effect of the restoring force, i.e., $\alpha=0$ in the last equation in (\ref{eq:bc-fl}).
It is found in \fref{fig:theory-numerical} (c) that there is a discrepancy between the analytical result (\ref{eq:amp-OF}) (dashed-dotted line) and our numerical result (dashed line) calculated by us on the basis of their model but with no approximations. 
According to the stability analysis of the ice-water interface in the papers (Ogawa and Furukawa 2002, Schewe and Riordon 2003), the instability of the ice-water interface occurs by the Laplace instability due to the thermal diffusion into the air. Moreover the flow in the thin water film makes the temperature distribution uniform, thus inhibiting the Laplace instability. They conclude that ripples of centimeter-scale wavelengths appear as a result of the competition between these two effects, and that the ripples on icicles should migrate downward. However, there are serious problems with this interpretation. 
First, our numerical calculation showed that even the length scale of ripples cannot be determined from their model because $\sigma_{*}^{(r)}>0$ for any wave number as shown by the dashed line in \fref{fig:theory-numerical} (c). 
This means that there exists no stabilization mechanism of the ice-water interface.
Second, according to the Laplace instability the latent heat is more rapidly lost from the convex surfaces than concave surfaces, resulting in faster ice growth on icicles's convex protrusions of the icicles than on its concave indentations. 
However, the Laplace instability cannot explain the translation mechanism of ripples.

Our numerical results indicate that the approximation used to derive (\ref{eq:amp-OF}) is obviously incorrect. The same faulty approximation was made when deriving (\ref{eq:amp-OF}) from our theoretical framework in (Ueno 2004). The comparison of the dashed line to dashed-dotted line in \fref{fig:theory-numerical} (c) shows that the $\mu^{2}$ term becomes dominant for $\mu>0.13$. Hence we cannot neglect the $\mu^{2}$ term in the equations and boundary conditions when deriving (\ref{eq:amp-OF}). In the case of (Ueno 2003), however, there exists already a stable region for $\mu<0.13$ as shown in \fref{fig:theory-numerical} (a), and thus the long wavelength approximation neglecting the higher order of $\mu$ is valid.
Even if we take into account the effect of $\alpha$ in the model (Ogawa and Furukawa 2002), the situation is not improved as shown by the solid lines in figures \ref{fig:theory-numerical} (c) and (d) as far as the boundary conditions in (\ref{eq:bc-Hl-OF}) are used.
The results are significantly different from figures \ref{fig:theory-numerical} (a) and (b): the system is unstable for perturbations of any wave number and the sign of phase velocity is opposite. 
The leading cause of these differences in the two models originates from the boundary conditions between (\ref{eq:bc-Hl-Ueno}) and (\ref{eq:bc-Hl-OF}) when solving (\ref{eq:gov-Hl}).

\section{Reconsideration of instability and stability of the ice-water interface \label{recon}}

\begin{figure}[t]
\begin{center}
\includegraphics[width=7.5cm,height=7.5cm,keepaspectratio,clip]{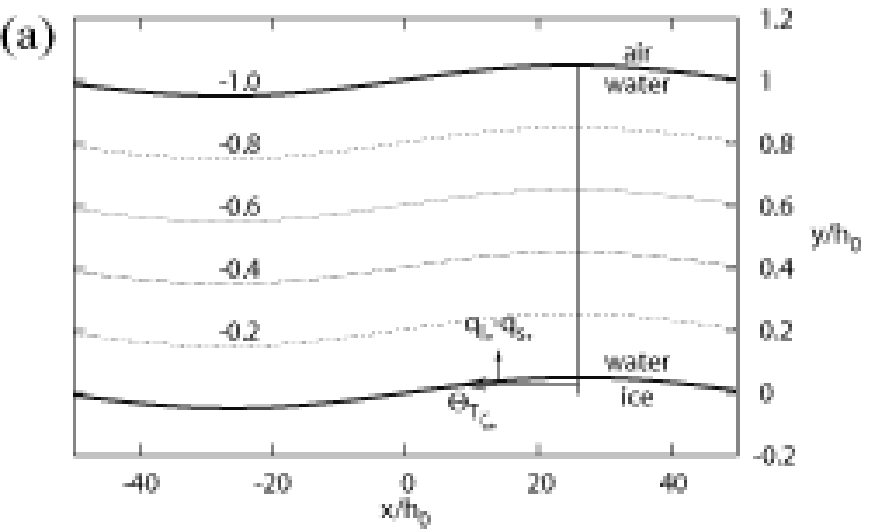}\hspace{3mm}
\includegraphics[width=7.5cm,height=7.5cm,keepaspectratio,clip]{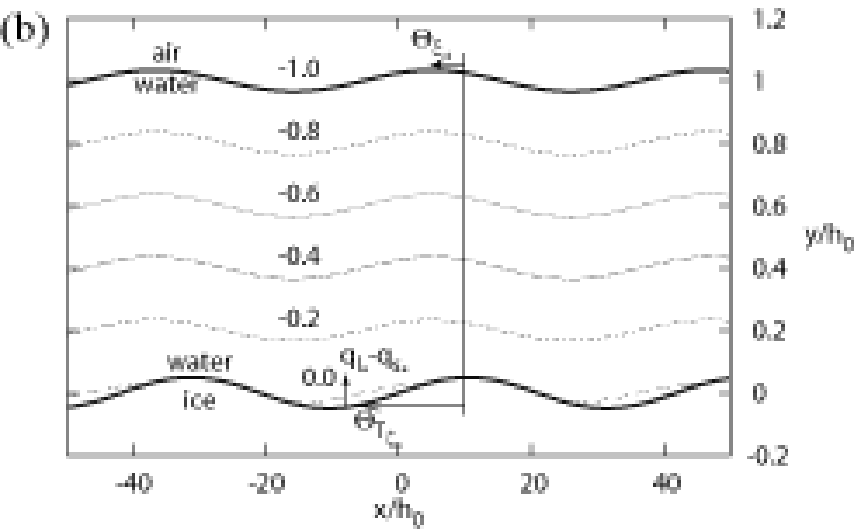}
\includegraphics[width=7.5cm,height=7.5cm,keepaspectratio,clip]{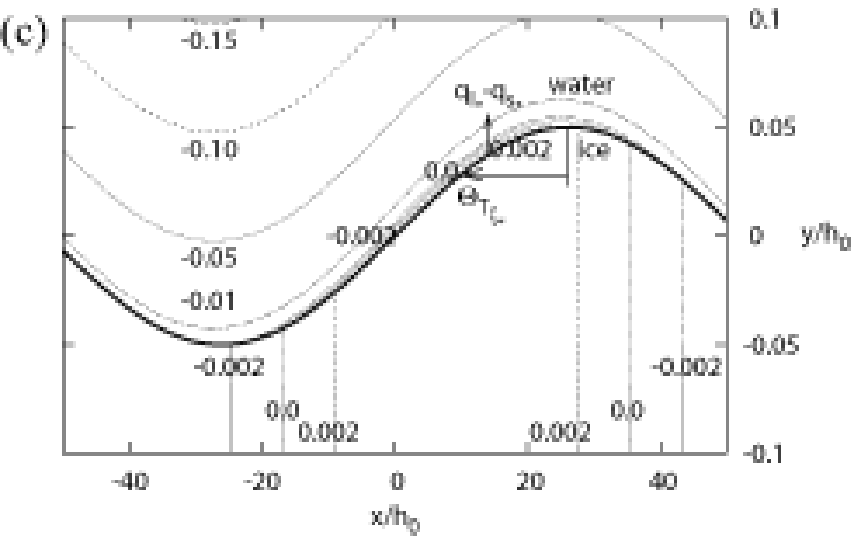}\hspace{3mm}
\includegraphics[width=7.5cm,height=7.5cm,keepaspectratio,clip]{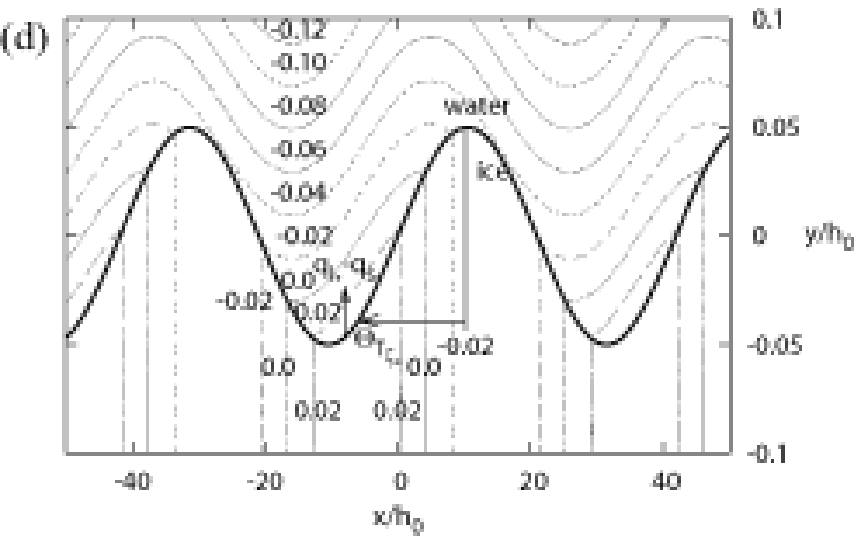}
\includegraphics[width=7.5cm,height=7.5cm,keepaspectratio,clip]{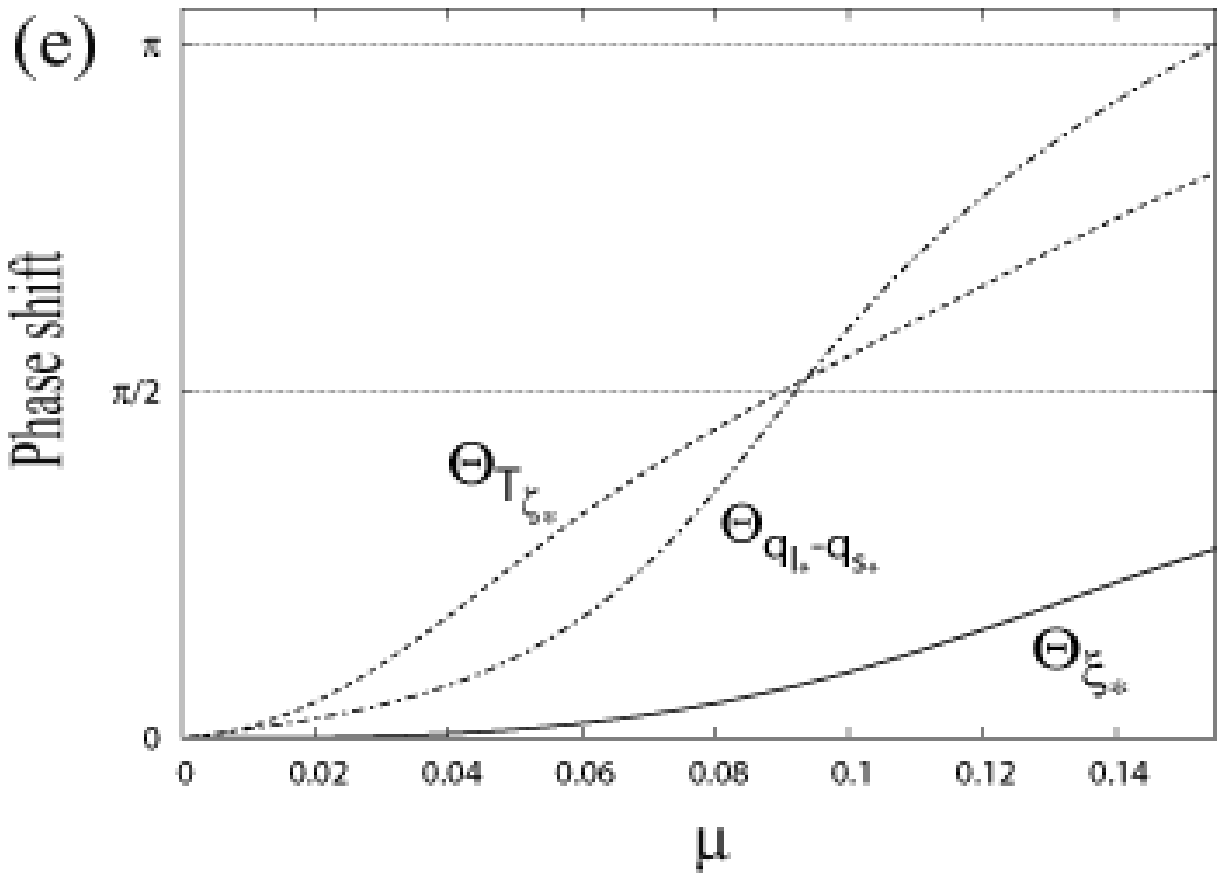}\hspace{3mm}
\includegraphics[width=7.5cm,height=7.5cm,keepaspectratio,clip]{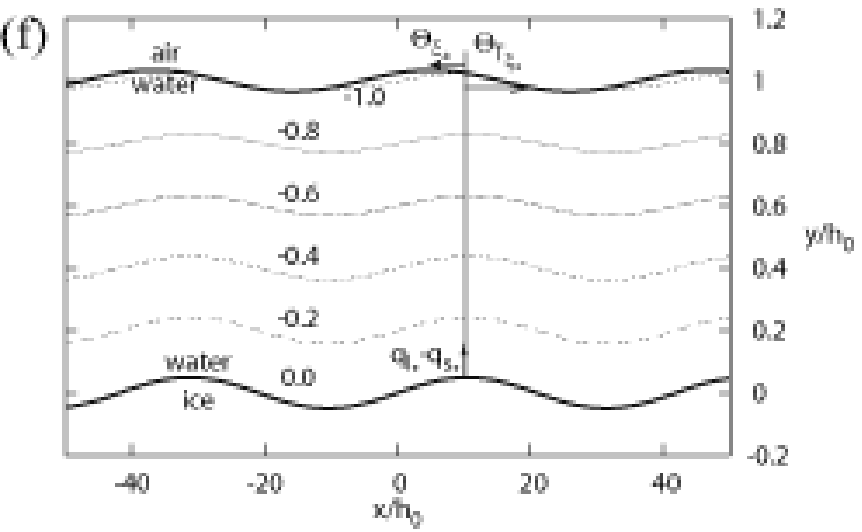}
\end{center}
\caption{(a)-(e) are in the case of $\Delta T_{sl*} \neq 0$ and $\Delta T_{la*}=0$. (f) is in the case of $\Delta T_{sl*}=0$ and $\Delta T_{la*} \neq 0$. (a) and (b) are isotherms in the water film at an unstable point ($\mu=0.06$, $\lambda=9.6$ mm) and a stable point ($\mu=0.15$, $\lambda=3.8$ mm) in \fref{fig:theory-numerical} (a), respectively, for $\delta_{b}=0.05$, $Q/l=50$ [(ml/h)/cm] and $\theta=\pi/2$.
(c) and (d) are isotherms in the water film in the vicinity of the ice-water interface in (a) and (b), respectively.
(e) $\Theta_{\xi_{*}}$ (solid line): phase shift of the water-air surface;
    $\Theta_{T_{\zeta_{*}}}$ (dashed line): phase shift of the temperature at the ice-water interface;
    $\Theta_{q_{l*}-q_{s*}}$ (dashed-dotted line): phase shift of total heat flux from the ice-water interface to the water and ice, against the ice-water interface. $\Theta_{q_{l*}-q_{s*}}=\pi/2$ at $\mu=0.092$, which 
corresponds to the point $\sigma_{*}^{(r)}=0$ of the solid line in \fref{fig:theory-numerical} (a).
(f) is isotherm in the water film at $\mu=0.15$ of the solid line in \fref{fig:theory-numerical} (c). $\Theta_{T_{\xi_{*}}}$: phase shift of the temperature at the water-air surface against the ice-water interface.}
\label{fig:isotherm}
\end{figure}

From the mathematical expression indicated by the terms including $\alpha$ with minus sign in (\ref{eq:amp-Ueno}), it was suggested that the restoring force due to gravity and surface tension is an important factor for the stabilization of the ice-water interface on a long length scale of about 1 cm (Ueno 2003). However, the detail of the morphological instability and/or stability mechanism of the ice-water interface was not clarified. 
In the subsequent paper (Ueno 2004), it was shown that there exists a phase shift between a disturbed ice-water interface and the maximum point of heat flux at its interface, and that the instability and/or stability of the interface is related to the magnitude of this phase shift.
However, the cause of the occurrence of such a phase shift was not well understood. Here, this is investigated in detail by drawing the isotherm in the water film. In figures \ref{fig:isotherm} (a), (b) and (f), the upper and lower solid lines are the water-air surface and the ice-water interface, respectively. Figures \ref{fig:isotherm} (c) and (d) show isotherms in the vicinity of the ice-water interface of figures \ref{fig:isotherm} (a) and (b), respectively.

It is convenient to express the temperature in the water film, 
$T_{l}=\bar{T}_{l}+T'_{l}=T_{sl}-\bar{G}_{l}y+H_{l}\bar{G}_{l}\zeta$, in the non-dimensional form, as follows:
\begin{equation}
T_{l*}(y_{*})\equiv \frac{T_{l}(y_{*})-T_{sl}}{T_{sl}-T_{la}}
=-y_{*}
+\delta_{b}H_{l}(y_{*}){\rm exp}[\sigma t+\rmi kx].
\label{eq:Tl}
\end{equation}
We also express the temperature in the ice, $T_{s}=\bar{T}_{s}+T'_{s}=T_{sl}+g_{s}(y){\rm exp}[\sigma t+\rmi kx]$, in the non-dimensional form. As far as $kb_{0} \gg 1$ ($b_{0} \gg 1.6$ mm for 1 cm ripple wavelength), (\ref{eq:sol-gs-Ueno}) can be approximated as follows: 
\begin{equation}
T_{s*}(y_{*}) \equiv \frac{T_{s}(y_{*})-T_{sl}}{T_{sl}-T_{la}}
=\delta_{b}{\rm exp}(\mu y_{*})(H_{l}|_{y_{*}=0}-1){\rm exp}[\sigma t+\rmi kx].
\label{eq:Ts}
\end{equation}
At the ice-water interface $y_{*}=\zeta_{*}$, taking the imaginary part for the perturbed part in (\ref{eq:Tl}) and (\ref{eq:Ts}) yields
\begin{equation}
\fl
T_{l*}|_{y_{*}=\zeta_{*}}=T_{s*}|_{y_{*}=\zeta_{*}}
=[(H_{l}^{(r)}|_{y_{*}=0}-1)^{2}+(H_{l}^{(i)}|_{y_{*}=0})^{2}]^{1/2}
\delta_{b}(t)\sin[k(x-v_{p}t)+\Theta_{T_{\zeta_{*}}}],
\label{eq:Tl-zeta}  
\end{equation}
where $\Theta_{T_{\zeta_{*}}}$ is a phase shift of the maximum point of temperature at $y_{*}=\zeta_{*}$ against that of the ice-water interface (see the horizontal arrows in the upstream direction in figures \ref{fig:isotherm} (a), (b), (c) and (d)). 
On the other hand, at the water-air surface $y_{*}=\xi_{*}$, taking the imaginary part for the perturbed part in (\ref{eq:Tl}) yields
\begin{eqnarray}
T_{l*}|_{y_{*}=\xi_{*}}
&=&-1+[(H_{l}^{(r)}|_{y_{*}=1}+f_{l}^{(r)}|_{y_{*}=1})^{2}
 +(H_{l}^{(i)}|_{y_{*}=1}+f_{la}^{(i)}|_{y_{*}=1})^{2}]^{1/2} \nonumber \\
&& \times \delta_{b}(t)\sin[k(x-v_{p}t)+\Theta_{T_{\xi_{*}}}], 
\label{eq:Tl-xi}
\end{eqnarray} 
where $\Theta_{T_{\xi_{*}}}$ is a phase shift of the maximum point of temperature at $y_{*}=\xi_{*}$ against that of the ice-water interface (see the horizontal arrow in the downstream direction in \fref{fig:isotherm} (f)). 

The conditions $T_{s}|_{y=\zeta}=T_{l}|_{y=\zeta}=T_{sl}+\Delta T_{sl}$ in (\ref{eq:Tb1-Ueno})
and
$T_{l}|_{y=\xi}=T_{a}|_{y=\xi}=T_{la}$ in (\ref{eq:Tb3-Ueno}) adopted in (Ueno 2003)
can be expressed in the non-dimensional form:
\begin{equation}
T_{s*}|_{y_{*}=\zeta_{*}}=T_{l*}|_{y_{*}=\zeta_{*}}=\Delta T_{sl*}, \qquad
T_{l*}|_{y_{*}=\xi_{*}}=T_{a*}|_{y_{*}=\xi_{*}}=-1. 
\label{eq:Ueno-Tsla}
\end{equation}
By comparing (\ref{eq:Ueno-Tsla}) to (\ref{eq:Tl-zeta}) and (\ref{eq:Tl-xi}), we obtain
\begin{eqnarray}
\fl
\Delta T_{sl*}=[(H_{l}^{(r)}|_{y_{*}=0}-1)^{2}+(H_{l}^{(i)}|_{y_{*}=0})^{2}]^{1/2}
\delta_{b}(t)\sin[k(x-v_{p}t)+\Theta_{T_{\zeta_{*}}}], \nonumber \\
H_{l}^{(r)}|_{y_{*}=1}=-f_{l}^{(r)}|_{y_{*}=1}, \qquad
H_{l}^{(i)}|_{y_{*}=1}=-f_{l}^{(i)}|_{y_{*}=1}.
\label{eq:DeltaTsl}
\end{eqnarray}
The last two equations in (\ref{eq:DeltaTsl}) are just the first equation in (\ref{eq:bc-Hl-Ueno}).
The dimensional form of the first equation in (\ref{eq:DeltaTsl}) is
$\Delta T_{sl}=(T_{sl}-T_{la})[(H_{l}^{(r)}|_{y_{*}=0}-1)^{2}+(H_{l}^{(i)}|_{y_{*}=0})^{2}]^{1/2}
\delta_{b}(t)\sin[k(x-v_{p}t)+\Theta_{T_{\zeta_{*}}}]$. 
Using the solution of $H_{l}$ at $Q/l=50$ [(ml/h)/cm] and $\theta=\pi/2$, for which $\sigma^{(r)}_{*}$ acquires a maximum value at $\mu=0.06$,  
the maximum value of $\Delta T_{sl}$ is $1.4 \times 10^{-4}$ $^{\circ}$C for the supercooling of $T_{sl}-T_{la}=0.03$ $^{\circ}$C of the water film (see Section \ref{exp}), $\mu=0.06$ and $\delta_{b}$=0.05. 
The temperature deviation from $T_{sl}=0$ $^{\circ}$C due to the Gibbs-Thomson effect evaluated at the same value of $\mu$ and $\delta_{b}$ is of the order of $10^{-6}$ $^{\circ}$C. 
Even if the value of $\Delta T_{sl}$ is extremely small but much greater than that due to the Gibbs Thomson effect, we cannot neglect the deviation because this contributes to the second terms in (\ref{eq:amp}), (\ref{eq:vp}), (\ref{eq:amp-Ueno}) and (\ref{eq:vp-Ueno}), which are represented by the dashed-dotted lines in figures \ref{fig:ampGsl-ampc-vpc} (b) and (c).

Figures \ref{fig:isotherm} (a) and (b) show isotherms in the water film obtained from (\ref{eq:Tl}) by using the solution $H_{l}$, determined by the boundary conditions in (\ref{eq:bc-Hl-Ueno}). It is found that the water-air surface is shifted by $\Theta_{\xi_{*}}$ in the upstream direction against the ice-water interface, and that $\Theta_{\xi_{*}}$ increases as $\mu$ increases, as shown by the solid line in \fref{fig:isotherm} (e). This phase shift is due to the effect of the restoring force in $f_{l}|_{y_{*}=1}$ in (\ref{eq:xi}).
The isotherm in the ice is determined by using (\ref{eq:Ts}). Since the typical value of the thickness of the water film is about 100 $\mu$m, figures \ref{fig:isotherm} (c) and (d) show isotherms around 10 $\mu$m from the ice-water interface. 

It should be noted that the isotherms in the water film are almost in phase with the shape of the water-air surface, as shown in figures \ref{fig:isotherm} (a) and (b). Since the water-air surface is shifted in the upstream direction against the ice-water interface, the temperature distribution become non-uniform in the vicinity of the ice-water interface. This non-uniformity does not disappear even at the ice-water interface, as a result, the temperature at the ice-water interface deviates by $\Delta T_{sl*}$. 
Figures \ref{fig:isotherm} (c) and (d) show that the maximum point of the temperature at the ice-water interface shifts by $\Theta_{T_{\zeta_{*}}}$ against that of the ice-water interface, which depends on $\mu$ as shown by the dashed line in \fref{fig:isotherm} (e). For example, \fref{fig:isotherm} (c) shows the isotherm near the ice-water interface with the wavelength of 9.6 mm. The deviation $\Delta T_{sl*}$ is positive on the upstream sides and is negative on the downstream sides of any protruded part of the ice-water interface. On the other hand, \fref{fig:isotherm} (d) shows the isotherm near the ice-water interface with the wavelength of 3.8 mm. The deviation $\Delta T_{sl*}$ is positive in any depressed region and is negative in any protruded region of the ice-water interface. 

We define the perturbed part of the non-dimensional heat flux from the ice-water interface to the water and from the ice to the ice-water interface, as 
$q_{l*}\equiv\Imag[-\partial T'_{l*}/\partial y_{*}|_{y_{*}=\zeta_{*}}]$ and
$q_{s*}\equiv\Imag[-K^{s}_{l}\partial T'_{s*}/\partial y_{*}|_{y_{*}=\zeta_{*}}]$, respectively,
where $T'_{l*}$ and $T'_{s*}$ represent the perturbed terms in (\ref{eq:Tl}) and (\ref{eq:Ts}).
Hence, the total heat flux from the ice-water interface to the water and ice can be expressed as follows:
\begin{eqnarray}
\fl
q_{l*}-q_{s*}
=\delta_{b}\Imag\left[\left\{-\frac{\rmd H_{l}}{\rmd y_{*}}\Big|_{y_{*}=0}
+K^{s}_{l}\mu(H_{l}|_{y_{*}=0}-1)\right\}{\rm exp}(\sigma t+\rmi kx)\right] \nonumber \\
=\left[\left\{-\frac{\rmd H_{l}^{(r)}}{\rmd y_{*}}\Big|_{y_{*}=0}
+K^{s}_{l}\mu(H_{l}^{(r)}|_{y_{*}=0}-1)\right\}^{2} \right. \nonumber \\
\left.       
+\left\{-\frac{\rmd H_{l}^{(i)}}{\rmd y_{*}}\Big|_{y_{*}=0}
+K^{s}_{l}\mu H_{l}^{(i)}|_{y_{*}=0}\right\}^{2}\right]^{1/2} 
\delta_{b}(t)\sin[k(x-v_{p}t)+\Theta_{q_{l*}-q_{s*}}],
\label{eq:ql-qs}
\end{eqnarray}
where $\Theta_{q_{l*}-q_{s*}}$ is a phase shift of the maximum point of the total heat flux at $y_{*}=\zeta_{*}$ against that of the ice-water interface and this changes as $\mu$ increases, as shown by the dashed-dotted line in \fref{fig:isotherm} (e). 
In order to avoid the temperature discontinuity by $\Delta T_{sl*}$ at the ice-water interface, 
the heat flux $q_{s*}$ by the thermal diffusion occurs. The perturbed heat flux in the ice exists only in the vicinity of the ice-water interface because it is observed from (\ref{eq:Ts}) that the non-uniformity of the temperature in the ice is exponentially attenuated far from the ice-water interface.

When $0 < \Theta_{q_{l*}-q_{s*}}<\pi/2$ as is the case in \fref{fig:isotherm} (c), the maximum point of $q_{l*}-q_{s*}$ shown by the vertical arrow is on the upstream side of any protruded part of the ice-water interface, which means that ice grows faster on the upstream side than on the downstream side of any protruded part of the ice-water interface. As a result, not only does the amplitude of perturbation grows, but ripples also move upward with time. 
On the other hand, when $\pi/2 < \Theta_{q_{l*}-q_{s*}}<\pi$ as in the case in \fref{fig:isotherm} (d), the maximum point of $q_{l*}-q_{s*}$ shown by the vertical arrow is in any depressed region of the ice-water interface. This means that ice grows faster at any depressed part of the ice-water interface, and grows slower at any protruded part. Accordingly, the disturbance of the ice-water interface diminishes with time and such a disturbance eventually cannot be observed. 
We find that a phase shift between a disturbed ice-water interface and the maximum point of heat flux at its interface
comes from the non-uniform temperature distribution at the ice-water interface due to the phase shift of the water-air surface against the ice-water interface. 
We also find that in order to explain the ripple migration it is necessary to cause an asymmetry in the temperature distribution between the upstream side and the downstream side of any protruded part of the ice-water interface.  

We define the characteristic time of shear rate as $\tau_{sh}$, which is just inverse of the shear rate $S$. The shear rate at the ice-water interface for the semi-parabolic shear flow $\bar{U}_{l}$ in (\ref{eq:sol-U-P}) is 
$S=\rmd \bar{U}_{l}/\rmd y|_{y=0}=2u_{0}/h_{0}$.
Hence, $\tau_{sh}=1/(2u_{0}/h_{0})=[3(g\sin\theta/\nu_{l})^{2}Q/l]^{-1/3}$ is of the order of $10^{-3}$ s for the typical range of $Q/l=10\sim 100$ [(ml/h)/cm] and $\theta=\pi/2$. We also define the thermal relaxation time $\tau_{a}\sim 1/(\kappa_{a}k^{2})$ and $\tau_{l}\sim 1/(\kappa_{l}k^{2})$ of fluctuations with wave number $k$, which are associated with the thermal diffusivity of the air, $\kappa_{a}=1.87\times10^{-5}$ $\rm m^{2}/s$, and that of water, $\kappa_{l}=1.3\times 10^{-7}$ $\rm m^{2}/s$, respectively ( see (Langer 1980, Caroli \etal 1992) for the relaxation time). For example, $\tau_{a}$ and $\tau_{l}$ are, respectively, of the order of 0.1 s and 10 s for about 1 cm ripple wavelength.
It is convenient to introduce a characteristic wave number $k_{c}$ by 
$\tau_{sh}=\tau_{l}$. When $\tau_{sh}<\tau_{l}$, temperature fluctuations with a wave number smaller than $k_{c}$ are affected by the shear flow before they can dissipate thermally.
On the other hand, when $\tau_{sh}>\tau_{l}$, temperature fluctuations with a wave number greater than $k_{c}$ can dissipate thermally without being affected by the shear flow (Onuki and Kawasaki 1979). 
In our system, the value of $k_{c}$ is about $10^{4.5}$, which corresponds to a wavelength of about 200 $\mu$m for the typical values of $u_{0} \sim 1$ cm/s and $h_{0} \sim 100$ $\mu$m. Therefore, the condition $\tau_{sh} \ll \tau_{l}$ is satisfied at the ice-water interface with a wavelength of about 1 cm. 

Based on the two time scales mentioned above, we will explain why did we choose the boundary condition $T_{l}|_{y=\xi}=T_{a}|_{y=\xi}=T_{la}$ at the water-air surface. 
We will also explain why the non-uniformity in the temperature distribution at the ice-water interface does not disappear, resulting in the temperature deviation $\Delta T_{sl}$ from $T_{sl}$.
A local temperature deviation of $\Delta T_{la}$ from $T_{la}$ at a disturbed water-air surface dissipates quickly by thermal diffusion in the air because shear stress is zero at the water-air surface. Therefore, it is reasonable to impose the boundary condition $T_{l}|_{y=\xi}=T_{a}|_{y=\xi}=T_{la}$ at the water-air surface. 
Since $\tau_{sh} \ll \tau_{a} \ll \tau_{l}$ for about 1 cm ripple wavelength, however,
the temperature distribution in the water film is determined so as to adapt the instantaneous disturbed shape of the water-air surface satisfying the boundary condition $T_{l}|_{y=\xi}=T_{a}|_{y=\xi}=T_{la}$ before a local temperature deviation $\Delta T_{sl}$ at the ice-water interface dissipates. This means that there is not enough time to relax the non-uniformity of the temperature at the ice-water interface thermally. 
That is why the temperature deviation $\Delta T_{sl}$ from $T_{sl}$ remains at the ice-water interface and the perturbed heat flux $q_{s*}$ by the thermal diffusion is maintained in the vicinity of the ice-water interface in the ice. 

On the other hand, the conditions $T_{s}|_{y=\zeta}=T_{l}|_{y=\zeta}=T_{sl}$ in (\ref{eq:Tb1-OF})
and $T_{l}|_{y=\xi}=T_{a}|_{y=\xi}=T_{la}+\Delta T_{la}$ in (\ref{eq:Tb3-OF}) adopted in (Ogawa and Furukawa 2002) 
can be expressed in the non-dimensional form: 
\begin{equation}
T_{s*}|_{y_{*}=\zeta_{*}}=T_{l*}|_{y_{*}=\zeta_{*}}=0, \qquad
T_{l*}|_{y_{*}=\xi_{*}}=T_{a*}|_{y_{*}=\xi_{*}}=-1+\Delta T_{la*}.
\label{eq:Ogawa-Tsla}
\end{equation}
By comparing (\ref{eq:Ogawa-Tsla}) to (\ref{eq:Tl-zeta}) and (\ref{eq:Tl-xi}), we obtain
\begin{eqnarray} 
H_{l}^{(r)}|_{y_{*}=0}=1, \qquad
H_{l}^{(i)}|_{y_{*}=0}=0, \nonumber \\
\fl 
\Delta T_{la*}=[(H_{l}^{(r)}|_{y_{*}=1}+f_{l}^{(r)}|_{y_{*}=1})^{2}
 +(H_{l}^{(i)}|_{y_{*}=1}+f_{l}^{(i)}|_{y_{*}=1})^{2}]^{1/2}
 \delta_{b}(t)\sin[k(x-v_{p}t)+\Theta_{T_{\xi_{*}}}]. \nonumber \\
\label{eq:DeltaTla}
\end{eqnarray}
The first two equations in (\ref{eq:DeltaTla}) are just the first equation in (\ref{eq:bc-Hl-OF}).
\Fref{fig:isotherm} (f) shows the isotherm in the water film obtained from (\ref{eq:Tl}) by using the solution $H_{l}$, determined by the boundary conditions in (\ref{eq:bc-Hl-OF}).
If we take into account the effect of the restoring force on the water-air surface in the model (Ogawa and Furukawa 2002), 
the water-air surface is shifted in the upstream direction against the ice-water interface as in \fref{fig:isotherm} (b). However, it should be noted that the isotherm in the water film is almost in phase with the shape of the ice-water interface, which is in contrast to the case in \fref{fig:isotherm} (b). The maximum point of heat flux $q_{l*}-q_{s*}$ indicated by the arrow is also different in between figures \ref{fig:isotherm} (b) and (f). In the case of \fref{fig:isotherm} (f), $q_{s*}=0$ because $\Delta T_{sl}=0$. 
Under such a situation, if there exists a phase shift between the water-air surface and the ice-water interface, the non-uniformity of the temperature occurs at the water-air surface, 
as shown in \fref{fig:isotherm} (f). The temperature at the water-air surface deviates by $\Delta T_{la}$ from $T_{la}$. 
The maximum point of the temperature at the water-air surface shifts by $\Theta_{T_{\xi_{*}}}$ against that of the ice-water interface. 
\Fref{fig:isotherm} (f) shows that $\Delta T_{la*}$ is negative on the upstream sides and $\Delta T_{la*}$ is positive on the downstream sides of any protruded part of the water-air surface.
The temperature distribution in the water film is determined so as to adapt the instantaneous disturbed shape of the ice-water interface satisfying the boundary condition $T_{s}|_{y=\zeta}=T_{l}|_{y=\zeta}=T_{sl}$ before a local temperature deviation $\Delta T_{la}$ at the water-air surface dissipates. 
However, this picture is physically inconsistent with the time scale $\tau_{sh} \ll \tau_{a} \ll \tau_{l}$ for the wavelength of interest. Actually, \fref{fig:theory-numerical} (c) obtained from the boundary condition $T_{s}|_{y=\zeta}=T_{l}|_{y=\zeta}=T_{sl}$ shows that ripples with a characteristic length scale cannot be observed on the ice surface because all modes are unstable.

The thermodynamics of fluids under shear flow is a challenging topic in modern non-equilibrium thermodynamics. 
For example, according to non-equilibrium molecular dynamics simulations of a system of spherical particles, the coexistence of crystal and shearing liquid flow cannot be accounted for by the equality of the chemical potentials of the crystal and liquid or by invoking a non-equilibrium analogue of the chemical potential (Butler and Harrowell 2002).
In our system, the ice-water interface is in a non-equilibrium state under the influence of the boundary of the water-air surface as indicated in (\ref{eq:bc-Hl-Ueno}) and shearing water flow. Since such a non-equilibrium contribution would be expected to change the water chemical potential,
there exists no physically reasonable definition of a non-equilibrium water chemical potential that would equal the chemical potential of the ice. Therefore, we did not impose the boundary condition $T_{s}|_{y=\zeta}=T_{l}|_{y=\zeta}=T_{sl}$ at the ice-water interface, where $T_{sl}$ is the equilibrium freezing temperature only when the chemical potential of water equals that of ice. 

\section{Experimental results \label{exp}}

\begin{figure}[t]
\begin{center}
\includegraphics[width=6cm,height=6cm,keepaspectratio,clip]{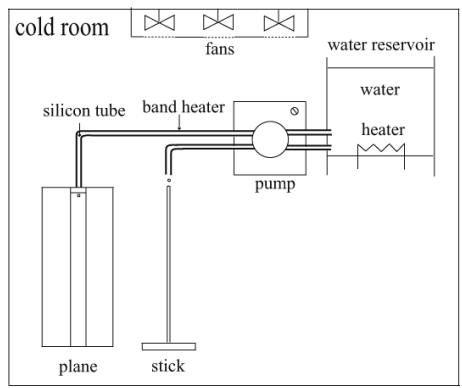}
\end{center}
\caption{Schematic view of apparatus set in a cold room. Water exits from the tip of the silicon tube covered with a band heater at the water supply rate of $Q$ ml/h and flows down along a wooden plane and a stick.}
\label{fig:apparatus}
\end{figure}

In this section, theoretical predictions are compared with experimental results. 
As shown in \fref{fig:apparatus}, we used a pump which can control the water supply rate within the range of 50 to 500 ml/h. Water was pumped from the reservoir and dripped from the tip of the silicon tube at the top of a gutter on an inclined plane and of a round stick. A wooden plane with $l_{x}=80$ cm in length, $l=3$ cm in width and 2 mm in thickness was inserted in the gutter of 2.5 cm in depth. The both sides and the back side of the gutter were covered with an insulation material to prevent the loss of latent heat at the sides. The stick was made of wood with dimensions of $l_{x}=80$ cm in length and 6 mm in diameter. The thermal conductivity of wood is normally around $0.10 \sim 0.15$ $\rm J/(m\,K\,s)$, which is much smaller than that of ice and aluminum. Since these instruments were set in a cold room, they were protected by a heating device in order to prevent the water from freezing in the silicon tube. The temperature of the water dripping from the top at the rate $Q$ ml/h was slightly above 0 $^{\circ}$C. The ceiling of the cold room was equipped with three fans. Although the fans were set to switch on and off periodically to make the temperature in the cold room uniform, large temperature fluctuations of $\pm 3$ $^{\circ}$C around $-9$ $^{\circ}$C were observed. The water reaches the supercooled state as it flows down along the plane and the stick. Ice grows from a portion of the supercooled water layer through which the latent heat of solidification is released into the ambient air below 0 $^{\circ}$C. The rest of the water drips from the lower edge of the plane and the stick. Our measurement of the mean growth rate $\bar{V}$ of the ice produced on the 6-mm diameter round stick was 1.7 mm/h, which was almost independent of the water supply rate. This result is consistent with previous theoretical (Makkonen 1988) and experimental results (Maeno \etal 1994). This is evident from (\ref{eq:V}) that $\bar{V}$ is independent of $Q/l$.   

\begin{figure}[t]
\begin{center}
\includegraphics[width=6cm,height=6cm,keepaspectratio,clip]{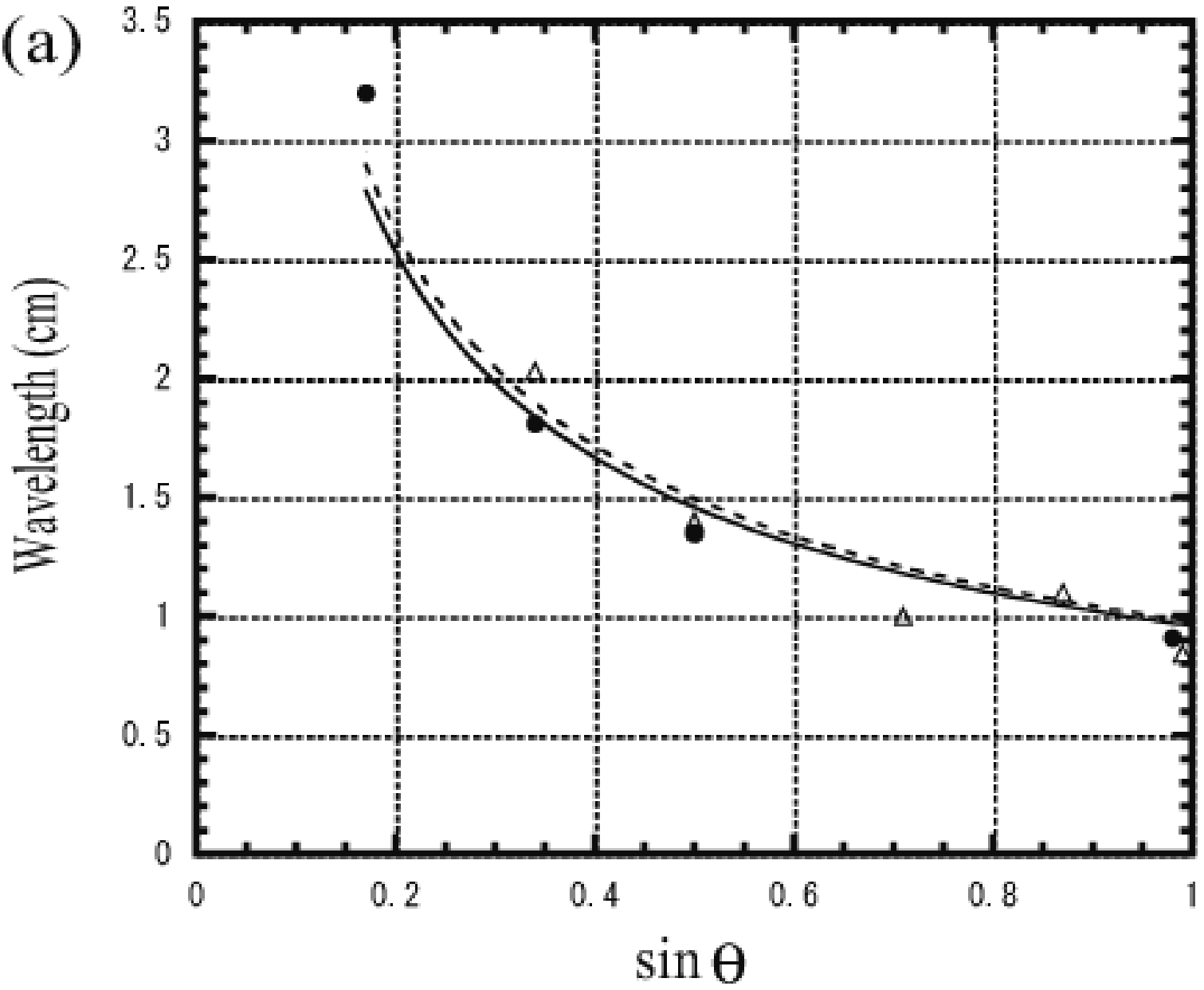}\hspace{5mm}
\hspace{0.2cm}
\includegraphics[width=6cm,height=6cm,keepaspectratio,clip]{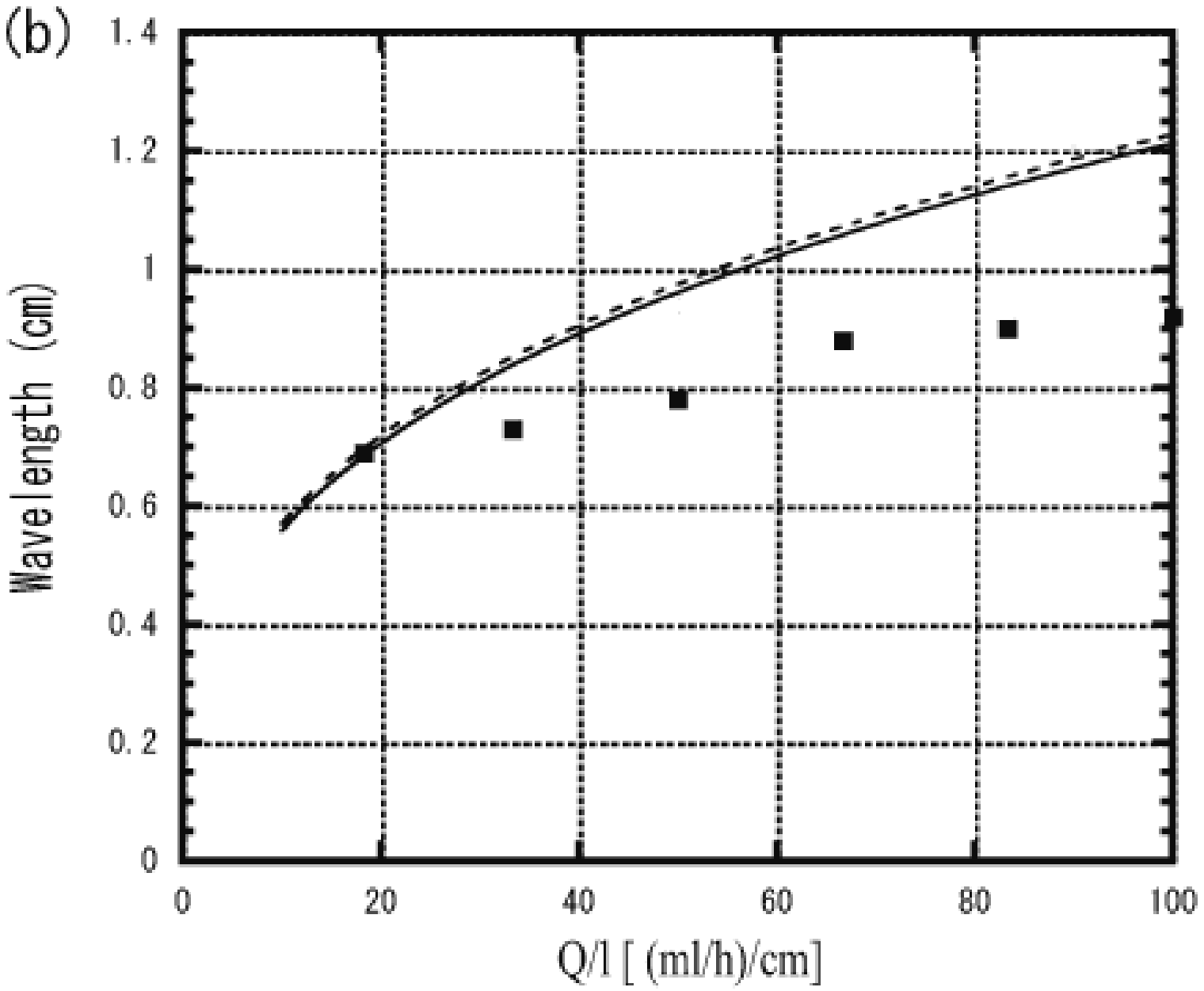}
\end{center}
\caption{Wavelength of ripples of ice produced on a gutter. (a) The wavelength versus $\sin\theta$ at $Q/l=$160/3 [(ml/h)/cm]. 
Solid line: numerical result; 
dashed line: analytical result (Ueno 2003);
$\opentriangle$: experimental result (Matsuda 1997); 
$\fullcircle$: our experimental result. 
(b) The wavelength versus $Q/l$ at $\theta=\pi/2$. 
Solid line: numerical result; 
dashed line:  analytical result (Ueno 2007);
$\fullsquare$: our experimental result.}
\label{fig:wavelength-theta-Qoverl}
\end{figure}

The wavelengths in figures \ref{fig:wavelength-theta-Qoverl} (a) and (b) are determined from the value of $\mu=kh_{0}$ at which $\sigma_{*}^{(r)}$ acquires a maximum value for a given $Q/l$ and $\theta$.
\Fref{fig:wavelength-theta-Qoverl} (a) shows the dependence of the ripple wavelength on the angle of the inclined plane at $Q/l=$160/3 [(ml/h)/cm]. As shown by the solid and dashed lines, our numerical and analytical results are in good agreement with the experimental results ($\opentriangle$ and $\fullcircle$). 
It is found that the wavelength of ripples increases with a decrease in angle.

Figure \ref{fig:wavelength-theta-Qoverl} (b) shows the dependence of the ripple wavelength on $Q/l$ at $\theta=\pi/2$. As shown by the solid and dashed lines, our numerical and analytical results show that the wavelength increases only gradually with an increase in $Q/l$. The experimental result $(\fullsquare)$
shows weaker dependence of the wavelength on $Q/l$ than that expected from the numerical and analytical results, but the qualitative behavior and order of wavelength are almost the same. 
It should be noted that a portion of the supplied water freezes, and that the rest flows down the surface of the ice. Therefore, $Q/l$ in $h_{0}$, $\Pec_{l}$ and $\Rey_{l}$ should be replaced by $Q/l-(\rho_{s}/\rho_{l})\bar{V}l_{x}$ from the mass conservation, where $\rho_{s}$ and $\rho_{l}$ are the density of ice and water, respectively. Using the value of $\rho_{s}/\rho_{l}=0.9$, $l_{x}=80$ cm and $\bar{V}=1.7$ mm/h under the assumption that the ice is completely produced along the gutter from the top to bottom, unfrozen water is given by $Q/l-12$ $\rm [(ml/h)/cm]$. Hence, the values of $h_{0}$, $\Pec_{l}$ and $\Rey_{l}$ are less than those estimated from $Q/l$ supplied from the top. Therefore, the actual ripple wavelength is expected to be slightly less than what the solid and dashed lines show in \fref{fig:wavelength-theta-Qoverl} (b).

\begin{figure}[t]
\begin{center}
\includegraphics[width=13cm,height=13cm,keepaspectratio,clip]{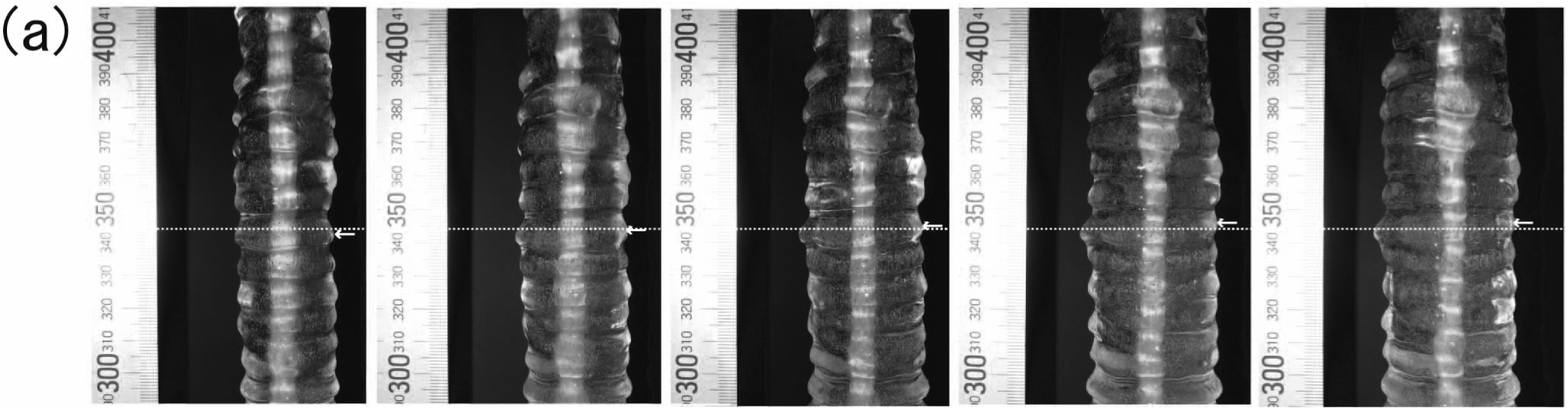}\\[0.5cm]
\includegraphics[width=13cm,height=13cm,keepaspectratio,clip]{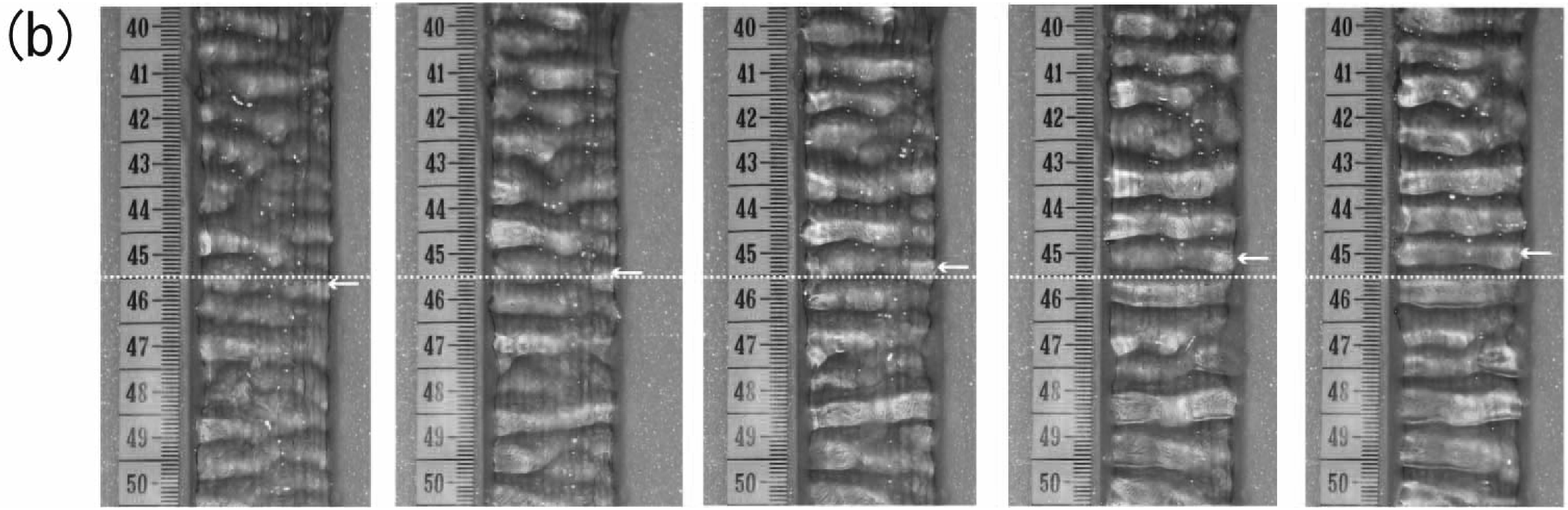}
\end{center}
\caption{A sequence of images showing upward movement of ripples of ice produced on (a) a 6-mm diameter round stick 
and (b) a gutter on a plane at $\theta=\pi/2$, after time 6, 7, 8, 9, and 10 hours (from left to right) at $Q=200$ ml/h. 
The mean displacement of ripples are (a) 3.2 mm and (b) 4.2 mm over 4 hours.}
\label{fig:ripple-movement}
\end{figure}

Our theory also predict that the ripples move upward with $|v_{p*}|=|v_{p}|/\bar{V}\approx 0.6$ at $Q=200$ $\rm ml/h$. This value is defined from $\mu=kh_{0}$ at which $\sigma_{*}^{(r)}$ acquires a maximum value for a given $Q/l$ and $\theta$.
It should be noted that the displacement of ripples depends on the growth rate $\bar{V}$. Using the measured mean value of $\bar{V}=1.7$ mm/h, we obtain $|v_{p}| \approx 1$ mm/h, meaning that the displacement over 4 hours is about 4 mm. Indeed, the observations in figures \ref{fig:ripple-movement} (a) and (b) show that all ripples move upward. For example, ripples indicated by the arrows pass through the dashed lines. Although all ripples move upward, their speeds are not uniform because some ripples sometimes do not move when some portion of the ice surface is not covered with water. The measured mean displacements for 4 hours in figures \ref{fig:ripple-movement} (a) and (b) are about 3.2 mm and 4.2 mm, respectively, which are of the same order as the theoretical results.
In our experiment, $Q$ from the top is kept constant. In the case of the ice produced on the round stick, the value of $Q/l$ decreases as ice grows because the value of $l$ increases with time $t$ as $2\pi(R_{0}+\bar{V}t)$ under the assumption that heat conduction into the wooden round stick through the ice is negligible, so that ice grows uniformly at $\bar{V}$, where $R_{0}$ is the stick radius. As a result, non-wetting parts on the ice surface increase as ice grows. The ripples produced initially almost disappeared over a 20-hour period due to sublimation. 
In this experiment, an upward movement of ripple was observed.
This result is consistent with the observation that many tiny air bubbles are trapped in the upstream region of any protruded part of an icicle, and line up in the upward direction during icicle growth (Maeno \etal 1994, Ueno 2007). 
 
In the absence of heat conduction into the substrate through the ice, from the second equation in (\ref{eq:bco(0)-sl-T}), we can estimate the degree of supercooling of the water film at $T_{sl}-T_{la}=L\bar{V}h_{0}/K_{l}=0.03$ $^{\circ}$C by using the measured value of $\bar{V}=1.7$ mm/h and the typical value of $h_{0}=100$ $\mu$m. 
Infrared instrumentation was used to keep the surface temperature of a thin water film flowing on growing ice below 0 $^{\circ}$C (Karev \etal 2007). The measurement showed that the surface temperature of the thin water film was always below 0 $^{\circ}$C. In our case too, it is necessary to measure the degree of supercooling of the water film accurately by such a non-destructive sensing technique. 

\section{Summary and Discussion \label{sum-dis}}

The validity of the approximations used in the two theoretical models (Ogawa and Furukawa 2002) and (Ueno 2003) was numerically investigated.
There was an apparent discrepancy for the amplification rate between the analytical result in (Ogawa and Furukawa 2002) and our numerical result, in spite of solving the same governing equations with the same boundary conditions as those used in the model (Ogawa and Furukawa 2002). The characteristic length scale of ripples could not be determined under their boundary conditions. 
On the other hand, the author's analytical results (Ueno 2003, Ueno 2007) were in good agreement with our numerical results, and the theoretical predictions were confirmed in our own experiments:
(i) the wavelength of ripples increases with a decrease in the angle of the inclined plane, 
(ii) the wavelength increases only gradually with an increase in the water supply rate per width, and 
(iii) the ripples move upward.

We also extended the theoretical framework (Ueno 2003) to include heat conduction into the substrate through the ice. If we take this into account, the wavelength depends on the ratio of the unperturbed temperature gradient in ice to that in water, $G^{s}_{l}$, which includes the parameters such as ice thickness, thermal boundary layer thickness in the air, as well as substrate and ambient air temperature. All wavelengths of ripples on icicles observed in nature as well as those produced experimentally on a wooden plane and stick have a centimeter-scale. The thermal conductivities of these materials are small, and so the heat conduction into the substrate can be neglected. Hence, we can say that the kind of universality of centimeter-scale ripples produced on an ice surface is due to $G^{s}_{l}=0$. 
On the other hand, in the case of $G^{s}_{l} \neq 0$, our stability analysis predicts a critical value of $G^{s}_{l}$, below which
the wavelength depends on the parameter $G^{s}_{l}$, but above which ripples do not appear. 
Our stability analysis would be applicable to prevent ripple formation on ice surfaces grown on a substrate for a given $Q/l$ and $\theta$. 
 
There is an analogy between wet growth and icicle growth (Makkonen and Lozowski 2008). In the case of wet growth, water is collected from the impingement of supercooled water droplets. Whereas, in the case of icicle growth, water is supplied from melting snow and ice at the root of the icicle. In both cases, the surface of ice is covered with  a supercooled water film, and ice grows from the portion of water film, by releasing latent heat into the ambient air below 0 $^{\circ}$C. 
From the point of view of engineering applications, the above analogy leads us to consider growth and morphology of ice on aircraft wings, wind turbine blades and aerial cables under wet icing condition, based on our theoretical framework. To do so, the present framework should be extended to include air flow, a supercooled water film motion driven by gravitational and aerodynamic forces, surface tension, and heat conduction through the ice into the object of cylindrical or arbitrary shape (Myers \etal 2002a, Myers \etal 2002b, Myers and Charpin 2004, Fu \etal 2006). In our model, the flow of water film is driven by gravity only, and both surface tension and gravity act on the water-air surface of the water film flowing down an inclined plane. The present basic velocity profile $\bar{U}_{l*}$ in the water film was derived from the free stress condition at the water-air surface. If an aerodynamic force also acts on the water-air surface, the basic profile of $\bar{U}_{l*}$ would change from the half-parabolic form. How were our results modified by the effect of the aerodynamic force?
The details will be discussed in a later paper.

\ack

This study was carried out within the framework of the NSERC/Hydro-Qu$\acute{\rm e}$bec/UQAC Industrial Chair on Atmospheric Icing of Power Network Equipment (CIGELE) and the Canada Research Chair on Engineering of Power Network Atmospheric Icing (INGIVRE) at the Universit$\acute{\rm e}$ du Qu$\acute{\rm e}$bec $\grave{\rm a}$ Chicoutimi. 
The authors would like to thank all CIGELE partners (Hydro-Qu$\acute{\rm e}$bec, Hydro One, R$\acute{\rm e}$seau Transport d'$\acute{\rm E}$lectricit$\acute{\rm e}$ (RTE) and $\acute{\rm E}$lectricit$\acute{\rm e}$ de France (EDF), Alcan Cable, K-Line Insulators, Tyco Electronics, Dual-ADE, and FUQAC) whose financial support made this research possible.
The authors would also like to thank S. Matsuda, P. Guba, and S. Goto for their useful comments. 

\section*{References}

\begin{harvard}

\item[]
Benjamin T. B 1957
Wave formation in laminar flow down an inclined plane
{\it J. Fluid Mech} {\bf 2} 554--74 

\item[]
Butler S and Harrowell P 2002
Factors determining crystal-liquid coexistence under shear
{\it Nature} {\bf 145} 1008--11 

\item[]
Caroli B, Caroli C and Roulet B 1992
Instabilities of planar solidification fronts
{\it Solids Far From Equilibrium},
ed Godr$\grave{\rm e}$che C (Cambridge: Cambridge University Press)

\item[]
Fu P, Farzaneh M and Bouchard G 2006
Two-dimensional modelling of the ice accretion process on transmission line wires and conductors
{\it Cold Reg. Sci. Technol} {\bf 46} 132--46 

\item[]
Karev A. R, Farzaneh M and Kollar L. E 2007
Measuring temperature of the ice surface during its formation by using infrared instrumentation
{\it Int. J. Heat Mass Transfer} {\bf 50} 566--79 

\item[]
Landau L. D and Lifschitz E. M 1959
{\it Fluid Mechanics}  
(London: Pergamon Press)

\item[]
Langer J. S 1980
Instabilities and pattern formation in crystal growth  
{\it Rev. Mod. Phys.} {\bf 52} 1--28

\item[]
Maeno N, Makkonen L, Nishimura K, Kosugi K and Takahashi T 1994
Growth rates of icicles  
{\it J.~Glaciol} {\bf 40} 319--26

\item[]
Makkonen L 1988
A model of icicle growth 
{\it J.~Glaciol} {\bf 34} 64--70

\item[]
Makkonen L and Lozowski E. P 2008
Numerical modelling of icing on power network equipment
\textit{Atmospheric Icing of Power Networks}, ed Farzaneh M (Berlin: Springer)

\item[]
Matsuda S 1997 
Experimental study on the wavy pattern of icicle surface 
Master's thesis, 
Institute of Low Temperature Science, Hokkaido University 

\item[]
Mullins W. W and Sekerka R. F 1963
Morphological stability of a particle growing by diffusion or heat flow
{\it J. Appl. Phys} {\bf 34} 323--29

\item[]
Myers T. G, Charpin J. P. F and Thompson C. P 2002a
Slowly accreting ice due to supercooled water impacting on a cold surface
{\it Phys Fluids} {\bf 14} 240--56 

\item[]
Myers T. G, Charpin J. P. F and Chapman  S. J 2002b
The flow and solidification of a thin fluid film on an arbitrary three-dimensional surface
{\it Phys Fluids} {\bf 14} 2788--803 

\item[]
Myers T. G and Charpin J. P. F 2004
A mathematical model for atmospheric ice accretion and water flow on a cold surface
{\it Int. J. Heat Mass Transfer} {\bf 47} 5483--500 
 
\item[]
Ogawa N and Furukawa Y 2002
Surface instability of icicles
{\it Phys. Rev. {\rm E}} {\bf 66} 041202

\item[]
Onuki A and Kawasaki K 1979
Nonequilibrium steady state of critical fluids under shear flow: a renormalization group approach
{\it Annals of Physics} {\bf 121} 456--528

\item[]
Oron A, Davis S. H and Bankoff S. G 1997
Long-scale evolution of thin liquid films 
{\it Rev. Mod. Phys} {\bf 69} 931--80

\item[]
Schewe P and Riordon J 2003
Icicle ripples
Physics Update, {\it Physics Today}, January p~9

\item[]
Short M. B, Baygents J. C and Goldstein R.E 2006
A free-boundary theory for the shape of the ideal dripping icicle
{\it Phys Fluids} {\bf 18} 083101

\item[]
Terada T 1947
{\it Collected Essays of Torahiko Terada} (Tokyo: Iwanami)

\item[]
Ueno K 2003
Pattern formation in crystal growth under parabolic shear flow 
{\it Phys. Rev. {\rm E}} {\bf 68} 021603

\item[]
Ueno K 2004 
Pattern formation in crystal growth under parabolic shear flow II  
{\it Phys. Rev. {\rm E}} {\bf 69} 051604

\item[]
Ueno K 2007
Characteristics of the wavelength of ripples on icicles
{\it Phys Fluids} {\bf 19} 093602

\end{harvard}

\end{document}